\pdfoutput=1
\documentclass[journal]{IEEEtran}
\usepackage{cite}

\ifCLASSINFOpdf
\else
\fi
\usepackage{amsmath}
\usepackage{array}

\usepackage{graphicx}      

\usepackage{amssymb}          
\usepackage{amsthm}
\usepackage{graphicx}         
\usepackage{subfigure}
\usepackage{color}
\usepackage{algorithm}
\usepackage{algorithmicx}
\usepackage{mathrsfs}
\usepackage{changes}
\usepackage{enumitem}
\usepackage{changebar}

\newtheorem{assumption}{Assumption}
\newtheorem{theorem}{Theorem}
\newtheorem{lem}{Lemma}
\newtheorem{rem}{Remark}

\begin{document}
%
\title{Fault Estimation Filter Design with Guaranteed Stability Using Markov Parameters}
%
%
%

\author{Yiming~Wan,~
        Tam\'{a}s~Keviczky,~
        and~Michel~Verhaegen
\thanks{This work has received funding from the European Union’s Seventh Framework Programme (FP7-RECONFIGURE/2007-2013) under grant agreement no. 314544.}
\thanks{Yiming Wan is with Massachusetts Institute of Technology, 77 Massachusetts Avenue, Cambridge, MA 02139. Email: ywan@mit.edu}
\thanks{Tam\'{a}s Keviczky and Michel Verhaegen are with Delft Center for Systems and Control, Delft University of Technology, 2628CD, The Netherlands. Emails: {t.keviczky, m.verhaegen}@tudelft.nl}}

%
%

\markboth{}%
{Shell \MakeLowercase{\textit{et al.}}: Bare Demo of IEEEtran.cls for IEEE Journals}
%



\maketitle

\begin{abstract}
For additive actuator and sensor faults, we propose a systematic method to design a state-space fault estimation filter directly from Markov parameters identified from fault-free data.
We address this problem by parameterizing a system-inversion-based fault estimation filter with the identified Markov parameters.
Even without building an explicit state-space plant model, our novel approach still allows the filter gain design for stabilization and suboptimal $\mathcal{H}_2$ performance. 
This design freedom cannot be achieved by other 
existing data-driven fault estimation filter designs so far.
Another benefit of our proposed design is the convenience of determining the state order: a higher state order of the filter leads to better estimation performance, at the cost of heavier computational burden. 
In contrast, order determination is cumbersome when using an identified state-space plant model for the filter design, because of the complicated propagation of the model mismatch into the fault estimation errors.
Simulations using an unstable aircraft system illustrate the effectiveness of the proposed new method.
\end{abstract}


%
\IEEEpeerreviewmaketitle

\section{Introduction}
\IEEEPARstart{O}{bserver}-based fault diagnosis techniques have been well established during the past two decades \cite{Ding2013book}. However, an explicit and accurate system model is often unknown in practice. In such situations, a conventional approach follows two steps: first identifying the state-space plant model from system input/output (I/O) data, and then designing observers for fault diagnosis \cite{Simani2003, Patward2005, Manuja2009, Mustata2014}. 
Different from the conventional two-step approach, the data-driven approach to fault diagnosis observer design has been investigated recently for additive sensor or actuator faults, without explicitly identifying a state-space plant model\cite{Ding2014book}.
As an alternative to multivariate statistical process monitoring such as principle component analysis (PCA) and partial least squares \cite{Qin2012, Gorin2015}, the data-driven fault diagnosis observer design offers a more powerful tool for highly dynamic systems, and allows developing systematic methods to address the same fault diagnosis performance criteria as the existing model-based approaches \cite{Ding2014book}.

Despite the recent progress in direct data-driven observer or filter design \cite{Novara2013}, research about dealing with completely unknown disturbances or faults in data-driven filter design has just started. 
Existing data-driven fault diagnosis observer design methods construct observers with either the parity vector \cite{Ding2009JPC} or the predictor Markov parameters (MPs) \cite{Dong2012a, Dong2012b} that can be identified from data. 
Data-driven fault estimation is much more involved than fault detection and isolation, because it is inherently related to inverting the underlying system whose model is unavailable. A non-recursive fault estimator was proposed in \cite{QinLi2001} and \cite{Lee2006} by minimizing the squared reconstructed prediction error in the residual subspace of a latent variable model. 
Recently, a receding horizon least-squares fault estimator has been proposed in \cite{WanKevicVerh2016} by using identified predictor MPs, which enables robust design that compensates for identification errors of the MPs.
Even fewer studies have been reported on data-driven design of recursive fault estimation observers or filters. 
Ding et al. first constructed a diagnostic observer realized with the identified parity vector, and then estimated faults as augmented state variables, see Chapter 10 of \cite{Ding2014book}. This augmented observer scheme, however, imposed certain limitations on how fault signals vary with time, thus introduced bias in fault estimates.
In contrast, without any assumptions on the dynamics of fault signals, Dong et al. first constructed a non-recursive fault estimation filter (FEF) in the form of finite impulse response (FIR) from the identified MPs, then used its state-space realization as a recursive FEF \cite{Dong2012c}.

As opposed to the model-based design, it is nontrivial to design a stable FEF directly from data without identifying an explicit state-space model. 
It is well known in model-based design that the existence of a stable inversion-based FEF is ensured when the fault subsystem has no unstable zeros \cite{Ants1978,Gill2007b,Kirt2011,Shi2014}. 
This property, however, cannot be guaranteed in current data-driven FEFs. For example, even under the above condition, 1) the parity vector based fault estimation observer in Chapter 10 of \cite{Ding2014book} needs the augmented fault state with assumed dynamics, which unnecessarily introduces estimation bias; and 2) the MP-based FEF in \cite{Dong2012c} might still be unstable. 

This paper focuses on data-driven design of FEF with stability guarantee, for additive actuator and sensor faults whose fault subsystem has arbitrary relative degrees. This problem is challenging, because it requires inverting the underlying plant dynamics without building an explicit state-space plant model. In order to pave the way for the data-driven design, an FEF is first constructed given the plant model in the predictor representation, which is structured into a residual generator and the inverse of the residual dynamics. 
Such a structured system-inversion-based FEF (SI-FEF) allows us to establish the link between the MPs of the SI-FEF and the predictor MPs.
By exploiting this link, our data-driven design method first computes the MPs of the SI-FEF with the predictor MPs identified from data, and then constructs a state-space realization of the SI-FEF from its MPs. 
Even without building an explicit state-space plant model, our data-driven design still allows the design freedom of the filter gain for stabilization and suboptimal $\mathcal{H}_2$ performance, which is missing in other existing data-driven designs.

Another benefit of our proposed design is related to the convenience of dertermining the state order: a higher state order of the filter leads to better estimation performance, at the cost of heavier computational burden. However, order determination is cumbersome when using an identified state-space plant model for the filter design, because of the complicated propagation of the model mismatch into the fault estimation errors.

This paper is organized as follows. In Section \ref{sect:prob}, we describe the system and formulate the data-driven FEF design problem. In Section \ref{sect:sspred_fest}, a SI-FEF is constructed given the plant model in the system predictor representation. Then the link between the MPs of the SI-FEF and the predictor MPs is established in Section \ref{sect:fefextend}. Our proposed data-driven design is developed in Section \ref{sect:design_using_Markov}. The advantages of this new method are illustrated via a numerical simulation example of an unstable aircraft system in Section \ref{sect:sim}. Finally, we give some concluding remarks in Section \ref{sect:conclusion}.

\section{Preliminaries and problem formulation}\label{sect:prob}
\subsection{Notations}
For the state-space model $\left( A, B, C, D \right)$, define Markov parameters as $H_0 = D$ and $H_i = C A^{i-1} B$ for $i>0$.
$\{ H_i \}$ represents the sequence of Markov parameters.
Let $\mathcal{O}_s$ and $\mathcal{T}_s$ denote the extended observability matrix with $s$ block elements and the lower triangular block-Toeplitz matrix with $s$ block columns and rows, respectively,  i.e.,
\begin{align}
& \mathcal{O}_{s} \left( A, C \right) = \left[ \begin{smallmatrix}
C \\
C A \\
\vdots \\
C A^{s-1}
\end{smallmatrix} \right],
\mathcal{T}_{s} \left( \{ H_i \} \right) = \left[ \begin{smallmatrix}
H_0 & 0 & \ldots & 0 \\
H_1 & H_0 & \ddots & \vdots \\
\vdots & \vdots & \ddots & 0  \\
H_{s-1} & H_{s-2} & \cdots & H_0
\end{smallmatrix} \right],
\label{eq:OL_TLmarkov} \\
& \text{or}\;\; \mathcal{T}_{s} \left( A, B, C, D \right) = \left[ \begin{smallmatrix}
D & 0 & \ldots & 0 \\
C B & D & \ddots & \vdots \\
\vdots & \vdots & \ddots & 0  \\
C A^{s -2} B & C A^{s -3} B & \cdots & D
\end{smallmatrix} \right].
\label{eq:TLu_ss}
\end{align}
Define 
\begin{equation}\label{eq:ukL}
\mathbf{y}_{k,l} = \left[ \begin{array}{ccc}
y^\mathrm{T}\left(k-l+1\right) & \cdots & y^\mathrm{T}\left(k\right)
\end{array} \right]^\mathrm{T}
\end{equation}
by stacking data vectors $\{y(i)\}$ in a sliding window $\left[ k-l+1, k \right]$.
$\text{diag} (X_1, X_2, \cdots, X_n)$ denotes a block-diagonal matrix.
$\mathbb{E}$ represents the mathematical expectation.

\subsection{System description}\label{sect:sys}
Consider a linear discrete-time system governed by
\begin{equation}\label{eq:sys_processform}
\begin{aligned}
\xi(k+1) &= A \xi(k) + Bu(k) + Ef(k) + w_1(k) \\
y(k) &= C \xi(k) + Du(k) + Gf(k) + w_2(k)
\end{aligned}
\end{equation}
where $\xi(k) \in \mathbb{R}^n$, $u(k) \in \mathbb{R}^{n_u}$, $y(k) \in \mathbb{R}^{n_y}$, and $f(k) \in \mathbb{R}^{n_f}$ represent the states, system inputs, output measurements, and latent faults, respectively. The process and measurement noises $w_1(k)$ and $w_2(k)$ are zero-mean white Gaussian.
$A, B, C, D, E, G$ are time-invariant matrices unavailable to the data-driven design. 
Assume that the system description (\ref{eq:sys_processform}) admits a Kalman filter for its fault-free subsystem, then this system (\ref{eq:sys_processform}) can be equivalently represented by the following Kalman predictor representation \cite{Mustata2014,WanKevicVerh2016}:
\begin{subequations}\label{eq:predictor}
	\begin{align}
	x(k+1) &= \Phi x(k) + \tilde B u(k) + \tilde E f(k) + K y(k), 
	\label{eq:predictor_dyn} \\
	y(k) &= C x(k) + D u(k) + G f(k) + e(k),
	\label{eq:predictor_out}
	\end{align}
\end{subequations}
where $x(k) \in \mathbb{R}^n$ and $e(k) \in \mathbb{R}^{n_y}$ are the predictor states and the innovation signal, respectively. $K$ is the steady-state Kalman gain, $\Phi = A - KC$, $\tilde B = B - KD$, and $\tilde E = E - KG$. No assumption is made about how the fault signals $f(k)$ evolve with time.

Define the MPs of the predictor representation (\ref{eq:predictor}) as
\begin{equation}\label{eq:markov_param}
\begin{aligned}
& H_i^u = \left\{ \begin{array}{ll}
D & i=0 \\
C \Phi^{i-1} \tilde B & i>0
\end{array} \right. , \;
H_i^y = \left\{ \begin{array}{ll}
0 & i=0 \\
C \Phi^{i-1} K & i>0
\end{array} \right. , \\
& H_i^f = \left\{ \begin{array}{ll}
G & i=0 \\
C \Phi^{i-1} \tilde E & i>0
\end{array} \right. .
\end{aligned}
\end{equation}
For the additive fault in the $j$th actuator or sensor, we may construct the predictor MPs $\{H_i^f\}$ from 
$\{H_i^u\}$ and $\{H_i^y\}$ as below according to (\ref{eq:predictor}) and (\ref{eq:markov_param}):
\begin{subequations}\label{eq:Hif}
\begin{align}
& j^\text{th} \text{ actuator fault: } \nonumber \\
& \quad E = B^{[j]}, G = D^{[j]}, H_i^f = (H_i^u)^{[j]} \quad i \geq 0; \\
& j^\text{th} \text{ sensor fault: } \nonumber \\
& \quad E = 0, G = I^{[j]}, 
H_i^f = \left\{ \begin{array}{ll}
I^{[j]} & i=0 \\
-(H_i^y)^{[j]} & 1<i\leq p
\end{array} \right. ,
\end{align}
\end{subequations}
where $X^{[j]}$ denotes the $j$th column of the matrix $X$.

The relative degree of the fault subsystem $(\Phi, \tilde E, C, G)$ can be determined from its MPs $\{ H_i^f \}$, i.e., the smallest nonnegative integer $\tau$ such that $H_\tau^f$ is nonzero.
Note that $\tau=0$ for sensor faults and $\tau > 0$ for actuator faults. 
We adopt the following assumption:

\begin{assumption}\label{ass:fault_rank}
	The $\tau$th MP of the fault subsystem $(\Phi, \tilde E, C, G)$ has full column rank, where $\tau$ is the relative degree of the fault subsystem.
\end{assumption}

Assumption \ref{ass:fault_rank} assumes sufficient number of measured outputs ($n_y \ge n_f$ for $H_\tau^f \in \mathbb{R}^{n_y \times n_f}$) and no collinearity among the fault directions to ensure the uniqueness of fault reconstruction. 
This assumption is common in fault estimation or input reconstruction literature, e.g., \cite{Dong2012a,Gill2007b, Marro2010, Kirt2011}.

\subsection{Markov parameter identification}\label{sect:ARXmdl}
When the accurate knowledge of the state-space description  (\ref{eq:sys_processform}) or  (\ref{eq:predictor}) is unavailable, we may identify the predictor MPs from data. It is well known that the predictor representation (\ref{eq:predictor}) can be approximated by the following vector ARX (VARX) model with arbitrary accuracy as the VARX order becomes sufficiently high \cite{Chiuso2007a, Veen2013}:
\begin{equation}\label{eqR:ARX}
\mathcal{A}(q^{-1})y(k) = \mathcal{B}(q^{-1}) u(k) + \mathcal{F}(q^{-1}) f(k) + v(k)
\end{equation}
where $q^{-1}$ is the backward shift operator, $\mathcal{A}(q^{-1}) = I - \sum\limits_{i=1}^{p} M_i^y q^{-i}$,
$\mathcal{B}(q^{-1}) = \sum\limits_{i=0}^{p} M_i^u q^{-i}$, $\mathcal{F}(q^{-1}) = \sum\limits_{i=0}^{p} M_i^f q^{-i}$, ${v(k)} \in \mathbb{R}^{n_y}$ represent the noise signal. 
Therefore, the coefficients of the high-order VARX approximation can be the estimates of the predictor MPs, i.e.,  
\begin{equation}\label{eq:MPest}
H_i^s \approx \left\{\begin{array}{ll}
M_i^s & 0 \leq i \leq p \\
0   & i > p
\end{array} \right.,
\text{ for } s \text{ represents } u, y, f.
\end{equation} 
Note that $H_0^y = 0$ is already known in (\ref{eq:markov_param}). For more detailed derivations, we refer to Section 2.2 of \cite{Veen2013}.

With the fault-free identification data, we can identify the VARX coefficients $\{M_i^u\}$ and $\{M_i^y\}$ as the estimates of the predictor MPs $\{H_i^u\}$ and $\{H_i^y\}$, and then construct $\{H_i^f\}$ for the additive faults according to (\ref{eq:Hif}).
The residual signal ${v(k)} = \mathcal{A}(q^{-1})y(k) - \mathcal{B}(q^{-1}) u(k)$ generated from the identification data approximates the innovation $e(k)$ of the predictor (\ref{eq:predictor}), and can be used to estimate the innovation covariance as
$$\Sigma_e = \text{cov}\left( \mathcal{A}(q^{-1})y(k) - \mathcal{B}(q^{-1}) u(k) \right).$$

In practice, data collected under faulty conditions may be seldom available, or  recorded without a reliable description of the fault type \cite{Ding2014book}. Therefore, no faulty historical data is used in our data-driven design.

\begin{rem}\label{rmk:p}
In theory, an infinite-order VARX model is needed to fully represent a Kalman predictor (\ref{eq:predictor}). Therefore, the identification of the predictor MPs actually requires identifying an infinite-order VARX model. For this purpose, we adopt a finite high-order VARX approximation. 
It should be noted that the consistent estimation of the infinite-order VARX coefficients, i.e., the predictor MPs, does not follow the conventional rules in the case of identifying a finite-order VARX model \cite{Kuer2005,Chiuso2007a}. 
In this paper, an empirical approach is used to select the order $p$ for the high-order VARX model:
first, we may identify a low-order VARX or VARMA model in order to roughly estimate the predictor poles and the noise variance; then, we determine the order $p$ according to the location of the predictor poles, so that the remaining predictor MPs can be well approximated by zero compared to the noise level.  
Note that the VARX order selection also involves a trade-off, i.e., selecting a higher order leads to smaller bias but larger variance in the identified predictor MPs.
\end{rem}

\subsection{Data-driven design of fault estimation filter}
\label{sect:prob_form_C}
Given the predictor MPs $\{ H_i^u, H_i^y, H_i^f \}$ identified offline from data as in Section \ref{sect:ARXmdl}, the basic idea of a system-inversion-based fault estimator follows two steps: 
\begin{enumerate}
	\item[(\romannumeral1)] Residual generation using the online I/O data, i.e., $r(k) = \mathcal{A}(q^{-1})y(k) - \mathcal{B}(q^{-1}) u(k)$. Then the residual dynamics is $r(k) = \mathcal{F}(q^{-1}) f(k) + e(k)$ according to (\ref{eqR:ARX}).	
	\item[(\romannumeral2)] $\tau$-delay fault estimation by processing the residual signal with the $\tau$-delay left inverse of $\mathcal{F}(q^{-1})$, i.e., $\hat f(k-\tau) = \mathcal{F}^{\mathrm{inv}}(q^{-1}) r(k)$, with $\mathcal{F}^{\mathrm{inv}}(q^{-1}) \mathcal{F}(q^{-1}) = q^{-\tau} I_{n_f}$.
\end{enumerate}
Note that the estimation delay $\tau$ in the above step (\romannumeral2) is due to the fact that the residual signal $r(k)$ contains only the fault information up to the time instant $k-\tau$ according to the definition of the relative degree $\tau$.
Finding a stable left inverse system $\mathcal{F}^{\mathrm{inv}}(q^{-1})$ is a long-studied problem in the literature \cite{Moylan1977, Ants1978, Hou1998, Xiong2003, Gill2007b, Marro2010, Kirt2011}. The capability of placing poles of the left inverse system is critical to the stability and performance of the system-inversion-based fault estimation. To achieve this capability, an explicit state-space plant model is needed in most system inversion literature, e.g., \cite{Moylan1977, Ants1978, Hou1998, Xiong2003, Marro2010}, Chapter 3 of \cite{Gill2007b}, and the references therein. Such a pole tuning or placement capability becomes non-trivial if only the knowledge of an input-output plant model is available, which is the case for the data-driven design problem in this paper.
This prevents the applicability of the data-driven methods in many situations.

In this subsection, we will briefly review the existing approaches for constructing the left inverse system $\mathcal{F}^{\mathrm{inv}}(q^{-1})$ from the predictor MPs $\{ H_i^u, H_i^y, H_i^f \}$, and point out their limitations that motivate our research.

One category of a fault estimator is in the form of an FIR filter. Note that the FIR filter is actually an approximated left inverse of $\mathcal{F}(q^{-1})$, and its construction avoids stability and pole placement mentioned above.
Assume the order of this FIR filter to be $L$, then the residuals $r(k-L+1), \cdots, r(k-1), r(k)$ are involved to produce the fault estimate at time instant $k$. By stacking the residual signal over the time window $[k-L+1,k]$, we obtain the stacked residual vector 
\begin{equation}\label{eqR:rkL}
\begin{aligned}
\mathbf{r}_{k,L} = \Psi_f \mathbf{f}_{k,L+p} 
+ \mathbf{e}_{k,L},
\end{aligned}
\end{equation}
according to the VARX model (\ref{eqR:ARX}), where $\mathbf{r}_{k,L}$, $\mathbf{f}_{k,L+p}$, and $\mathbf{e}_{k,L}$ are defined as in (\ref{eq:ukL}), and  
\begin{equation*}
\Psi_f = \left[ \begin{smallmatrix}
M_p^f & M_{p-1}^f & \cdots & M_1^f & M_0^f & 0 & \cdots & 0 \\
M_{p+1}^f & M_{p}^f & \cdots & M_2^f & M_1^f & M_0^f & \ddots & \vdots \\
\vdots & \vdots & \ddots & \vdots & \vdots & \vdots & \ddots & 0 \\
M_{p+L-1}^f & M_{p+L-2}^f & \cdots & M_{L}^f & M_{L-1}^f & M_{L-2}^f & \cdots & M_{0}^f
\end{smallmatrix} \right].
\end{equation*}
Since the coefficients $M_i^f$ of the high-order VARX model approximate the predictor MPs $\{H_i^f\}$, the first $p$ block-columns of $\Psi_f$ approximate the Hankel matrix of the fault subsystem $(\Phi, \tilde E, C, G)$. Because of this link, the first $p$ block-columns of $\Psi_f$ can be ill-conditioned or rank deficient for large $L$ and $p$, see Theorem 6.1 of \cite{Kata2005}. 
This implies that the stacked fault vector $\mathbf{f}_{k,L+p}$ may not be uniquely or reliably reconstructed from the residual vector $\mathbf{r}_{k,L}$. 
Similarly to this above reason, the dynamic PCA based fault reconstruction in \cite{QinLi2001} might be non-unique, and it estimates a linear combination of the true faults, as can be seen from the equation (24) of \cite{QinLi2001}. Such a fault estimate is obviously biased.
Instead of reconstructing the entire fault vector $\mathbf{f}_{k,L+p}$, recent research in \cite{Dong2012c, WanKevicVerh2016} produced the $\tau$-delay fault estimate $f(k-\tau)$ by applying an FIR filter on the residual signal. 
It was shown in \cite{WanKevicVerh2016} that unbiased fault estimates can be achieved asymptotically as the FIR filter order $L$ goes to infinity, if the fault subsystem has no unstable zeros. 

The second category of a fault estimator is in the state-space form for the benefit of efficient online recursive computation.
The conventional approach designs a state-space FEF based on a state-space plant model identified from data.
This approach might lead to large fault estimation errors,
because there is no effective method in literature to suppress the complicated nonlinear propagation of the state-space system identification errors into the fault estimates. 
Another approach in the recent data-driven design is to construct the FEF as the state-space realization of the aforementioned FIR filter \cite{Dong2012c}. 
However, such an obtained state-space FEF is not guaranteed to be stable, and its poles cannot be tuned in the design. 

\begin{figure}[!h]
	\begin{center}
		\includegraphics[width=7.5cm]{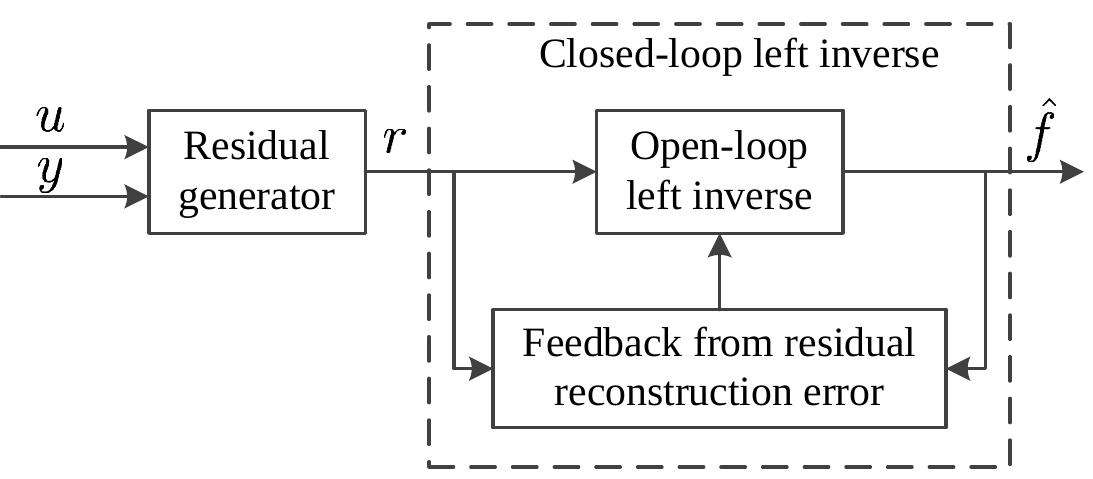}    
		\caption{Our proposed fault estimation filter scheme}
		\label{fig:FEF_scheme}
	\end{center}
\end{figure}

\begin{figure}[!h]
	\begin{center}
		\includegraphics[width=8.7cm]{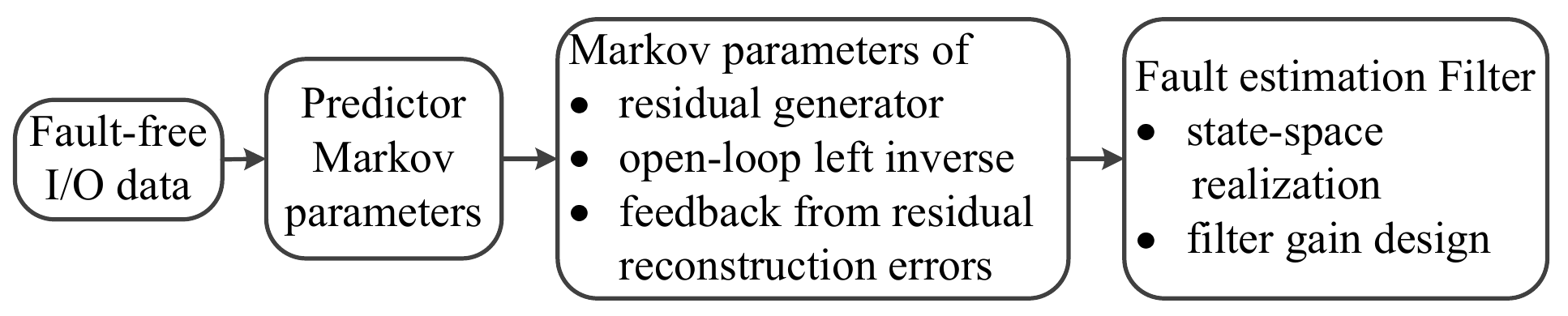}    
		\caption{Basic idea of our proposed data-driven design}
		\label{fig:d2idea}
	\end{center}
\end{figure}

The aim of this paper is to construct a state-space SI-FEF with tunable stable poles by using the predictor MPs $\{ H_i^u, H_i^y, H_i^f \}$ identified from data. As summarized in Figure \ref{fig:FEF_scheme}, our proposed approach constructs the SI-FEF from a residual generator, an open-loop left inverse of the residual dynamics, and the feedback from the residual reconstruction errors. This structure allows (\romannumeral1)
establishing the link connecting $\{ H_i^u, H_i^y, H_i^f \}$ and the MPs of the SI-FEF as in Figure \ref{fig:d2idea}, and (\romannumeral2) designing the feedback gain of the residual reconstruction errors for stability and performance.

Note that the identification errors of the predictor MPs affect the fault estimation performance. This issue has been investigated recently in \cite{WanKevicVerh2016} for the robust data-driven design of a receding horizon fault estimator. How to address the same issue for a data-driven state-space FEF can be investigated only after the stability is ensured. In this paper, we focus on the stability guarantee, and leave the robustness issue for future research.
Because of this reason, the identification errors of the predictor MPs are not explicitly considered in this paper, and we will use the notations $\{H_i^u\}$, $\{H_i^y\}$, $\{H_i^f\}$ for both the true predictor MPs and their estimates.

\section{System-inversion-based fault estimation filter using the predictor representation}
\label{sect:sspred_fest}
As the foundation for our data-driven design, we construct an SI-FEF in this section by exploiting the accurate knowledge of the predictor representation (\ref{eq:predictor}). 
Note that all system matrices in (\ref{eq:predictor}) are unknown to our data-driven design, but used here for establishing the link between the predictor MPs and the MPs of the SI-FEF in Section \ref{sect:fefextend}.

Firstly, we decompose the predictor (\ref{eq:predictor}) into two subsystems: 
\begin{subequations}\label{eq:res_gen_ss}
	\begin{align}
	x_1 (k+1) &= \Phi x_1 (k) + \tilde B u(k) + K y(k) \label{eq:res_gen_ss1}\\
	y_1(k) &= C x_1 (k) + D u(k), \; x_1(0) = \hat x(0), 
	\label{eq:res_gen_ss2}
	\end{align}
\end{subequations}
and 
\begin{subequations}\label{eq:res_dyn_ss}
	\begin{align}
	x_2 (k+1) &= \Phi x_2 (k) + \tilde E f(k) \label{eq:res_dyn_1} \\
	r(k) &= C x_2 (k) + G f(k) + e(k), \label{eq:res_dyn_2}
	\end{align}
\end{subequations}
such that $x(k) = x_1(k) + x_2(k)$ and $y(k) = y_1(k) + r(k)$.
Starting from an initial guess of the predictor state $\hat x(0)$, the subsystem (\ref{eq:res_gen_ss}) predicts the output  without accounting for the fault. As shown in Figure \ref{fig:FEF_scheme}, (\ref{eq:res_gen_ss}) is used to generate a residual signal $r(k) = y(k) - y_1(k)$ from the I/O data. Then, the subsystem (\ref{eq:res_dyn_ss}) is the residual dynamics decoupled from the I/O data. This will be used in the following to design a closed-loop left inverse system as depicted in Figure \ref{fig:FEF_scheme}.

Since the fault subsystem $(\Phi, \tilde E, C, G)$ has the relative degree $\tau$ (see Assumption \ref{ass:fault_rank}), 
the residual signal at the time instant $k+\tau$ is needed to produce a fault estimate $\hat f(k)$, 
which introduces an estimation delay when $\tau > 0$.
Considering this estimation delay, we construct
the following equation $r(k+\tau)$ by successively substituting (\ref{eq:res_gen_ss1}) and (\ref{eq:res_dyn_1}) into (\ref{eq:res_gen_ss2}) and (\ref{eq:res_dyn_2}), respectively:
\begin{subequations}\label{eq:res_ext_output}
	\begin{align}
	{r}(k+\tau) &= y(k+\tau) - y_1(k+\tau) \nonumber \\
	&= - C \Phi^\tau x_1 (k) - {B}_{\tau+1}^u \mathbf{u}_{k+\tau, \tau+1} \nonumber \\
	&\quad + {B}_{\tau+1}^y \mathbf{y}_{k+\tau, \tau+1}  \label{eq:res_ext_output_1} \\
	&= C \Phi^\tau  x_2 (k) + H_{\tau}^f f(k) + {e}(k+\tau) \label{eq:res_ext_output_3}
	\end{align}
\end{subequations}
where $\mathbf{u}_{k+\tau, \tau+1}$ and $\mathbf{y}_{k+\tau, \tau+1}$ are defined in (\ref{eq:ukL}), 
${B}_{\tau+1}^{u}$ and ${B}_{\tau+1}^{y}$ are respectively defined as 
\begin{subequations}
	\begin{align}
	&{B}_{\tau+1}^{u} = \left[ \begin{matrix}
	H_{\tau}^u & H_{\tau-1}^u & \cdots & H_0^u
	\end{matrix} \right], \nonumber \\
	&{B}_{\tau+1}^{y} = \left[ \begin{matrix}
	-H_{\tau}^y & -H_{\tau-1}^y & \cdots & -H_1^y & I
	\end{matrix} \right]. \nonumber
	\end{align}
\end{subequations}
In (\ref{eq:res_ext_output_3}), we use the fact $H_i^f = 0$ for $i<\tau$ according to the definition of relative degree $\tau$. 

From (\ref{eq:res_ext_output_3}), $f(k)$ can be estimated as below by using $x_2(k)$ and a left inverse matrix $\Pi$ of $H_\tau^f$:
\begin{equation}\label{eq:Pi_matrix}
\hat f(k) = \Pi \left[ {r}(k+\tau) - C \Phi^{\tau} x_2(k) \right], \,
\Pi H_\tau^f = I.
\end{equation}
The left inverse matrix $\Pi$ is a design parameter, whose existence is ensured by Assumption \ref{ass:fault_rank}.
Since the state $x_2(k)$ is actually unknown,  we construct the following left inverse of the residual dynamics (\ref{eq:res_dyn_ss}) and (\ref{eq:res_ext_output}) in the state-space form which jointly estimates the state and the fault:
\begin{subequations}\label{eqr:leftinv}
	\begin{align}
	\hat x_2 (k+1) &= \Phi \hat x_2(k) + \tilde E \hat f(k) +  K_r \tilde r(k+\tau) \label{eqr:leftinv_dyn} \\
	\hat f(k) &= \Pi \left[ {r}(k+\tau) - C \Phi^{\tau} \hat x_2(k) \right] \label{eqr:leftinv_fest} \\
	\tilde r(k+\tau) &= {r}(k+\tau)- {\hat r}(k+\tau) \label{eqr:leftinv_rhat}\\
	&= {r}(k+\tau) - C \Phi^{\tau} \hat x_2(k) - {H}_{\tau}^f \hat f(k). \label{eqr:leftinv_rtilde}
	\end{align}
\end{subequations}
By replacing the state $x_2$ and the fault $f$ with their estimates $\hat x_2$ and $\hat f$, $\hat r(k+\tau) = C \Phi^\tau \hat x_2(k) + H_\tau^f \hat f(k)$ in (\ref{eqr:leftinv_rhat}) and (\ref{eqr:leftinv_rtilde}) follows (\ref{eq:res_ext_output_3}) to reconstruct the residual signal from the state and fault estimates. Then $\tilde r(k+\tau) = r(k+\tau) - \hat r(k+\tau)$ is the residual reconstruction error.
(\ref{eqr:leftinv_dyn}) is a copy of the residual dynamics (\ref{eq:res_dyn_1}) with a feedback term $K_r \tilde r(k+\tau)$ from the residual reconstruction error $\tilde r(k+\tau)$. Similarly, (\ref{eqr:leftinv_fest}) constructs the fault estimate $\hat f(k)$ by following (\ref{eq:Pi_matrix}).   
By substituting (\ref{eqr:leftinv_fest}) into (\ref{eqr:leftinv_dyn}) and (\ref{eqr:leftinv_rtilde}), respectively, the left inverse (\ref{eqr:leftinv}) can be equivalently rewritten as
\begin{subequations}\label{eqr:leftinv_1}
	\begin{align}
	\hat x_2 (k+1) &= \Phi_1 \hat x_2(k) + B_1 r(k+\tau) + K_r \tilde r(k+\tau) \label{eqr:leftinv_dyn_2}\\
	\hat f(k) &= C_1 \hat x_2(k) + D_1 {r}(k+\tau) \label{eqr:leftinv_fest_2} \\
	\tilde r(k+\tau) &= -C_2 \hat x_2(k) + D_2 {r}(k+\tau) \label{eqr:leftinv_rtilde_2}
	\end{align}
\end{subequations}
with 
\begin{gather}
\Phi_1 = \Phi - \tilde E \Pi C \Phi^{\tau}, \;
B_1 = \tilde E \Pi, \; C_1 = - \Pi C \Phi^{\tau}, \;  \label{eq:Phi1_B1} \\
D_1 = \Pi,\; C_2 = (I - {H}_{\tau}^f \Pi) C \Phi^\tau, \;
D_2 = I - {H}_{\tau}^f \Pi. \label{eq:C2_D2}
\end{gather}

With $K_r = 0$, $(\Phi_1, B_1, C_1, D_1)$ in the above left inverse system  is referred to as an open-loop left inverse.
With the feedback gain $K_r$, the residual reconstruction error $\tilde {r}(k+\tau)$ is used as a feedback signal to stabilize the above left inverse. 
Such a structured form of the closed-loop inverse (\ref{eqr:leftinv_1}), i.e., the combination of the open-loop left inverse and the feedback from the residual reconstruction errors $\tilde r(k+\tau)$, enables our data-driven design in Sections \ref{sect:fefextend} and \ref{sect:design_using_Markov}.

By cascading the residual generator (\ref{eq:res_gen_ss}) and the left inverse (\ref{eqr:leftinv_1}), we obtain the SI-FEF as below:
\begin{equation}\label{eq:sfestfilter_reduced}
\begin{aligned}
\hat x (k+1) &= \Phi_1 \hat x (k) + {B}_f \mathbf{u}_{k+\tau,\tau+1} + {K}_f \mathbf{y}_{k+\tau,\tau+1} \\
&\quad + K_r  {\tilde r}(k+\tau) \\
\hat f (k) &= C_1 \hat x(k) + D_{f,1} \mathbf{u}_{k+\tau,\tau+1} + G_{f,1} \mathbf{y}_{k+\tau,\tau+1}, \\
{\tilde r}(k+\tau) &= - C_2 \hat x (k) - D_{f,2} \mathbf{u}_{k+\tau,\tau+1} + G_{f,2} \mathbf{y}_{k+\tau,\tau+1}.
\end{aligned}
\end{equation}
Note that $\hat x(k) = x_1(k) + \hat x_2(k)$ is an estimate of the predictor state $x(k)$, because $\hat x_2(k)$ is the estimate of $x_2(k)$ and the predictor state is decomposed as $x(k) = x_1(k) + x_2(k)$. 
In the above SI-FEF, $\Phi_1$, $B_1$, $C_1$, $D_1$, $C_2$ and $D_2$ are defined in (\ref{eq:Phi1_B1}) and (\ref{eq:C2_D2}), respectively, and
\begin{equation}\label{eq:BKtau}
\tilde B_{\tau} = \left[\begin{matrix}
\tilde{B} & \mathbf{0}_{n_x \times \tau n_u }
\end{matrix}\right],\;
K_\tau = \left[\begin{matrix}
K & \mathbf{0}_{n_x \times \tau n_y }
\end{matrix}\right],
\end{equation}
\begin{equation*}\label{eq:BDKGf}
\begin{array}{rlrl}
B_f &= \tilde B_\tau - B_1 B_{\tau+1}^u,  &
K_f &= K_\tau + B_1 B_{\tau+1}^y,  \\
D_{f,1} &= - {D_1} B_{\tau+1}^u,  &
G_{f,1} &= {D_1} B_{\tau+1}^y, \\
D_{f,2} &=  D_2 B_{\tau+1}^u, &
G_{f,2} &= D_2 B_{\tau+1}^y.  
\end{array}
\end{equation*}

Next, the error dynamics of the SI-FEF (\ref{eq:sfestfilter_reduced}) is analyzed by defining the state estimation error $\tilde x (k) = x (k) - \hat x(k)$ and the fault estimation error $\tilde f(k) = f(k) - \hat f(k)$: 
\begin{equation}\label{eq:fest_err_dyn}
\begin{aligned}
\tilde x (k+1) &= \left( \Phi_1 - K_r C_2 \right) \tilde x (k) - \left( B_1 + K_r D_2 \right) {e}(k+\tau) \\
\tilde f(k) &= C_1 \tilde x(k) - D_1 {e}(k+\tau).
\end{aligned}
\end{equation}
Therefore, if the pair $\left( \Phi_1, C_2 \right)$ is observable or detectable, there exists a stabilizing gain $K_r$ in (\ref{eq:fest_err_dyn}), such that starting from any arbitrary initial estimate $\hat x(0)$, unbiasedness of the estimates $\hat x(k)$ and $\hat f(k)$ can be achieved asymptotically, i.e., 
$\mathop {\lim }\limits_{k \rightarrow \infty} \mathbb{E} \left(\tilde x(k)\right) = 0$ and 
$\mathop {\lim }\limits_{k \rightarrow \infty} \mathbb{E} \left(\tilde f(k)\right) = 0$.

\begin{theorem}\label{thm:stabilizability}
	$( \Phi_1, C_2 )$ is observable if the fault subsystem $( \Phi, \tilde E, C \Phi^\tau, {H}_{\tau}^f )$ has no invariant zeros;
	$( \Phi_1, C_2 )$ is detectable if all invariant zeros of $( \Phi, \tilde E, C \Phi^\tau, {H}_{\tau}^f )$ are stable. 
\end{theorem}

The proof is given in the Appendix. 
Theorem \ref{thm:stabilizability} shows how the observability or detectability of the pair $\left( \Phi_1, C_2 \right)$ is determined by the invariant zeros of the underlying fault subsystem. Thus it provides a sufficient condition for the existence of a stabilizing filter gain for the SI-FEF (\ref{eq:sfestfilter_reduced}).

The SI-FEF (\ref{eq:sfestfilter_reduced}) produces both the state estimate $\hat x(k)$ and the fault estimate $\hat f(k)$. However, it is different from the simultaneous state and input estimation filter proposed by \cite{Gill2007, Gill2007a} in two aspects: (\romannumeral1) the condition in Theorem \ref{thm:stabilizability} that ensures stabilization and asymptotic unbiasedness was not provided; (\romannumeral2) only the special cases of $\tau=0$ and $\tau=1$ were discussed in \cite{Gill2007, Gill2007a}.

How to design $\Pi$ in (\ref{eq:Pi_matrix}) and $K_r$ in (\ref{eq:fest_err_dyn}) for our data-driven design problem will be discussed in Section \ref{sect:gain_design}.

\section{Markov parameters of system-inversion-based fault estimation filter}\label{sect:fefextend}
As illustrated in Figure \ref{fig:d2idea}, after the MPs of the SI-FEF (\ref{eq:sfestfilter_reduced}) are computed, the state-space realization of the SI-FEF can be constructed. 
In this section, we establish the link for computing MPs of the SI-FEF (\ref{eq:sfestfilter_reduced}) from the predictor MPs $\{H_i^u, H_i^y, H_i^f\}$.

As the first step towards the above goal, we rewrite the residual generator (\ref{eq:res_gen_ss}), the left inverse system (\ref{eqr:leftinv_1}), and the SI-FEF (\ref{eq:sfestfilter_reduced}) into extended forms over a time window.
With $k_0 = k-L+1$, we define 
\begin{align}
\mathbf{\bar z}_{k,L} = \left[ \begin{array}{ccc}
\mathbf{z}_{k_0+\tau,\tau+1}^\mathrm{T} & \cdots & \mathbf{z}_{k+\tau,\tau+1}^\mathrm{T}
\end{array} \right]^\mathrm{T}, \label{eq:barskL} 
\end{align}
by stacking $\mathbf{z}_{k+\tau,\tau+1} = \left[ \begin{matrix}
\mathbf{u}_{k+\tau, \tau+1}^\text{T} &
\mathbf{y}_{k+\tau, \tau+1}^\text{T}
\end{matrix} \right]^\text{T}$ over the time window $[k_0,k]$.
According to (\ref{eq:res_gen_ss1}), (\ref{eq:res_dyn_1}), and (\ref{eq:res_ext_output}), the stacked residual vector $\mathbf{r}_{k+\tau,L}$ over the time window $\left[ k_0, k \right]$ can be written in the extended form
\begin{subequations}\label{eq:resL_compute}
	\begin{align}
	\mathbf{r}_{k+\tau,L} &= \mathcal{O}_L \left( \Phi, - C \Phi^{\tau} \right) \, x_1(k_0) + \mathscr{T}_L^z \mathbf{\bar z}_{k,L} \\
	&= \mathcal{O}_L \left( \Phi, C \Phi^{\tau} \right) \,  x_2 (k_0) + \mathscr{T}_L^f \mathbf{f}_{k,L} + \mathbf{e}_{k+\tau,L} \label{eq:resL_compute2}
	\end{align}
\end{subequations}
with $\tilde B_\tau$ and $K_\tau$ defined in (\ref{eq:BKtau}),
\begin{align}
\mathscr{T}_L^z &= \mathcal{T}_L \left( \Phi, \left[\begin{matrix}
\tilde B_\tau & K_\tau
\end{matrix} \right], - C \Phi^{\tau},
\left[\begin{matrix}
-{B}_{\tau+1}^u & {B}_{\tau+1}^y
\end{matrix}\right] \right), \label{eq:scrTLz} \\
\mathscr{T}_L^f &= \mathcal{T}_L \left( \Phi, \tilde E,  C \Phi^{\tau}, {H}_{\tau}^f \right). \label{eq:scrTLf} 
\end{align}

Since the residual generator (\ref{eq:res_gen_ss}) has the initial state $x_1(k_0) = \hat x(k_0)$, the closed-loop left inverse (\ref{eqr:leftinv_1}) then has the initial state $\hat x_2(k_0) = 0$ according to $\hat x(k) = x_1(k) + \hat x_2(k)$ in (\ref{eq:sfestfilter_reduced}). 
Hence, the closed-loop left inverse (\ref{eqr:leftinv_1}) can be transformed into the following extended form over the time window $\left[ k_0, k \right]$ to produce the stacked fault estimates $\mathbf{\hat f}_{k,L}$: 
\begin{equation}\label{eq:festkL_inv}
\begin{aligned}
\mathbf{\hat f}_{k,L} = \mathcal{K}_L \mathbf{r}_{k+\tau,L} = \left( \mathcal{G}_L + \mathcal{M}_L \mathcal{J}_L \right)
\mathbf{r}_{k+\tau,L},
\end{aligned}
\end{equation}
with
\begin{subequations}\label{eq:GL_JL}
	\begin{align}
	\mathcal{K}_L &= \mathcal{T}_L \left( \Phi_1 - K_r C_2, B_1 + K_r D_2, C_1, D_1 \right), \label{eq:KL} \\
	\mathcal{G}_L &= \mathcal{T}_L \left( \Phi_1, B_1, C_1, D_1 \right),  \label{eq:GL}\\
	\mathcal{J}_L &= I - \mathscr{T}_L^f \mathcal{G}_L = \mathcal{T}_L \left( \Phi_1, B_1, -C_2, D_2 \right), \label{eq:JL} \\
	\mathcal{M}_L &= \mathcal{T}_L \left( \Phi_1 - K_r C_2, K_r, C_1, 0 \right). \label{eq:ML}
	\end{align}
\end{subequations}
The proof of $\mathcal{K}_L = \mathcal{G}_L + \mathcal{M}_L \mathcal{J}_L$ in (\ref{eq:festkL_inv}) is given in the Appendix. Note that $\mathcal{K}_L$, $\mathcal{G}_L$, $\mathcal{J}_L$ and $\mathcal{M}_L$ are all block-Toeplitz matrices, and can be explained as below:
\begin{enumerate}[label=(\roman*)]
	\item $\mathcal{G}_L$ corresponds to the open-loop left inverse, i.e., (\ref{eqr:leftinv_1}) with $K_r = 0$;
	\item $\mathcal{J}_L$ produces the residual reconstruction errors $\tilde {r}(k+\tau)$ in (\ref{eqr:leftinv_1}) with $K_r = 0$;
	\item $\mathcal{M}_L$ corresponds to the feedback dynamics from the \\residual reconstruction errors $\tilde {r}(k+\tau)$ in the closed-loop inverse (\ref{eqr:leftinv_1}). 
\end{enumerate}

By substituting the residual generator (\ref{eq:resL_compute}) into the extended closed-loop inverse (\ref{eq:festkL_inv}), the following extended form of the SI-FEF (\ref{eq:sfestfilter_reduced}) is obtained:
\begin{subequations}\label{eq:festfilter_extended0}
	\begin{align}
	{{\mathbf{\hat {{f}}}}_{k,L}} &= \mathcal{O}_L \left( \Phi_1 - K_r C_2, C_1 \right) \, x_1(k_0) +
	\left( \mathcal{R}_L + \mathcal{M}_L \mathcal{Q}_L \right)
	\mathbf{\bar z}_{k,L} \label{eq:festfilter_extended} \\
	&= \mathcal{O}_L \left( \Phi_1 - K_r C_2, -C_1 \right) \, x_2 (k_0) + \mathbf{f}_{k,L}
	+ \mathcal{K}_L \mathbf{e}_{k+\tau,L} \label{eq:festfilter_extended_dyn}
	\end{align}
\end{subequations}
with
\begin{subequations}\label{eq:RL_QL}
	\begin{align}
	\mathcal{R}_L &= \mathcal{G}_L \mathscr{T}_L^z
	= \mathcal{T}_L \left( \Phi_1, [\begin{matrix} B_f & K_f \end{matrix}], C_1,
	[\begin{matrix} D_{f,1} & G_{f,1} \end{matrix}] \right), \label{eq:RL}\\
	\mathcal{Q}_L &= \mathcal{J}_L \mathscr{T}_L^z
	= \mathcal{T}_L \left( \Phi_1, [\begin{matrix} B_f & K_f \end{matrix}], -C_2,
	[\begin{matrix} -D_{f,2} & G_{f,2} \end{matrix}] \right).  \label{eq:QL}   
	\end{align}
\end{subequations}
Similarly to $\mathcal{G}_L$ and $\mathcal{J}_L$ in (\ref{eq:festkL_inv}), 
$\mathcal{R}_L$ and $\mathcal{Q}_L$ correspond to two subsystems of the SI-FEF (\ref{eq:sfestfilter_reduced}) with $K_r = 0$, which produce $\hat f(k)$ and $\tilde r(k+\tau)$ in the open loop, respectively. $\mathcal{M}_L$ is the same feedback dynamics as in (\ref{eq:ML}).

The extended form (\ref{eq:festfilter_extended}) can be regarded as a batch estimator which provides the estimate $\mathbf{\hat f}_{k,L}$ from the I/O data $\mathbf{\bar z}_{k,L}$ and the initial state $x_1(k_0) = \hat x(k_0)$. Moreover, it can be seen from (\ref{eq:festfilter_extended_dyn}) that $\mathbf{\hat f}_{k,L}$ is a biased estimate of $\mathbf{f}_{k,L}$ due to the presence of unknown initial state $x_2(k_0)$. However, it follows from the definition of $\mathcal{O}_L \left( \Phi_1 - K_r C_2, -C_1 \right)$ in (\ref{eq:OL_TLmarkov}) that
\begin{equation*}
\begin{aligned}
\mathbb{E} \left( \hat f(k) - f(k) \right) = -C_1 \left(\Phi_1 - K_r C_2\right)^{L-1} x_2 (k_0),
\end{aligned}
\end{equation*}
where $\hat f(k)$ and $f(k)$ are the last $n_f$ entries of $\mathbf{\hat f}_{k,L}$ and $\mathbf{f}_{k,L}$, respectively.
The above equation shows that $\hat f(k)$, extracted from $\mathbf{\hat f}_{k,L}$ in (\ref{eq:festfilter_extended}), gives asymptotically unbiased fault estimation as $L$ goes to infinity, if $\Phi_1 - K_r C_2$ is stabilized given the condition in Theorem \ref{thm:stabilizability}.

In the above derivations, the block-Toeplitz matrices $\mathscr{T}_L^z$, $\mathscr{T}_L^f$, $\mathcal{\mathcal{G}}_L$, $\mathcal{\mathcal{J}}_L$, and $\mathcal{Q}_L$ are expressed with state-space matrices. For the data-driven design, the next step is to construct their corresponding MPs defined as
\begin{equation}\label{eq:RL_QL_Markov}
\begin{array}{c}
\mathscr{T}_L^z = \mathcal{T}_L \left( \{\mathscr{H}_i^z\} \right),\;
\mathscr{T}_L^f = \mathcal{T}_L \left( \{\mathscr{H}_i^f\} \right), 
\mathcal{\mathcal{G}}_L = \mathcal{T}_L \left( \{ G_i \} \right),\\
\mathcal{\mathcal{J}}_L = \mathcal{T}_L \left( \{ J_i \} \right), \;
\mathcal{R}_L = \mathcal{T}_L \left( \{ R_i \} \right), \; 
\mathcal{Q}_L = \mathcal{T}_L \left( \{ Q_i \} \right),
\end{array}
\end{equation}
from the predictor MPs $\{ H_i^u, H_i^y, H_i^f \}$.
To achieve this goal, we first need to take a closer look at $\mathscr{T}_L^z$,
$\mathscr{T}_L^f$ and $\mathcal{G}_L$ which are needed in computing $\mathcal{R}_L$ and $\mathcal{Q}_L$.
According to (\ref{eq:scrTLz}) and (\ref{eq:scrTLf}), the MPs 
$\{\mathscr{H}_i^z\}$ and $\{\mathscr{H}_i^f\}$ can be computed from the predictor MPs $\{ H_i^u, H_i^y, H_i^f \}$ as below:
\begin{align}
&\left\{
\begin{array}{l}
\begin{aligned}
\mathscr{H}_0^z &=
\left[\begin{matrix}
-B_{\tau+1}^u & B_{\tau+1}^y
\end{matrix}\right] \\ 
&= \left[\begin{matrix}
-H_{\tau}^u & \cdots & -H_0^u & -H_{\tau}^y & \cdots & -H_1^y & I
\end{matrix}\right]
\end{aligned} \\
\begin{aligned}
\mathscr{H}_i^z &=
- C \Phi^{\tau+i-1} \left[\begin{matrix}
\tilde B_\tau & K_\tau
\end{matrix} \right] \\
&= - \left[ \begin{matrix}
H_{\tau+i}^u & \mathbf{0}_{n_y \times \tau n_u} & H_{\tau+i}^y & \mathbf{0}_{n_y \times \tau n_y } 
\end{matrix} \right],\\ & \text{ for } 1 \le i \le L-1,
\end{aligned} 
\end{array}
\right. \label{eq:scrHz}\\
& \left\{
\begin{array}{l}
\mathscr{H}_0^f = {H}_{\tau}^f,\\
\begin{aligned}
\mathscr{H}_i^f = C \Phi^{\tau+i-1} \tilde E 
= H_{\tau+i}^f,\text{ for } 1 \le i \le L-1.
\end{aligned}
\end{array}
\right. \label{eq:scrHf}
\end{align}
As pointed out in the explanations below (\ref{eq:festkL_inv}) and (\ref{eq:GL_JL}), $\mathcal{G}_L$ is a left inverse matrix with block-Toeplitz structure for $\mathscr{T}_L^f$. Such a left inverse matrix is non-unique, but can be computed from the MPs $\{\mathscr{H}_i^f\}$. With $\Pi \mathscr{H}_0^f = \Pi H_\tau^f = I$ according to (\ref{eq:Pi_matrix}) and (\ref{eq:scrHf}),  one possible solution of $\mathcal{G}_L$ is given below:
\begin{equation}\label{eq:Gi}
\left\{
\begin{array}{l}
G_0 = \Pi, \\
G_i = - \sum\limits_{j=1}^i G_{i-j} \mathscr{H}_j^f G_0, \text{ for }1 \le i \le L-1.
\end{array}
\right.
\end{equation} 
which ensures $\mathcal{G}_L \mathscr{T}_L^f = I$.
Then, according to (\ref{eq:RL_QL}), the MPs of $\mathcal{R}_L$ can be computed as the convolution of $\{ G_i \}$ in (\ref{eq:Gi}) and $\{ \mathscr{H}_i^z \}$ in (\ref{eq:scrHz}):
\begin{equation}\label{eq:Ri}
R_i = \sum_{j=0}^i  G_{i-j} \mathscr{H}_{j}^z, \text{ for } 0 \leq i \leq L-1.
\end{equation}
Similarly, the MPs $\{ J_i, Q_i \}$ of $\mathcal{J}_L$ in (\ref{eq:JL}) and $\mathcal{Q}_L$ in (\ref{eq:RL_QL}) can be computed as
\begin{align}
&\left\{ \begin{array}{ll}
J_0 = I - \mathscr{H}_{0}^f G_{0},   \\
J_i = - \sum_{j=0}^i  \mathscr{H}_{i-j}^f G_{j}, \text{ for } 1 \leq i \leq L-1,  
\end{array} \right.  \label{eq:Ji}\\
&Q_i = \sum_{j=0}^i  J_{i-j} \mathscr{H}_{j}^z, \text{ for } 0 \le i \le L-1. \label{eq:Qi}
\end{align} 

\begin{figure}
	\begin{center}
		\includegraphics[width=8.5cm]{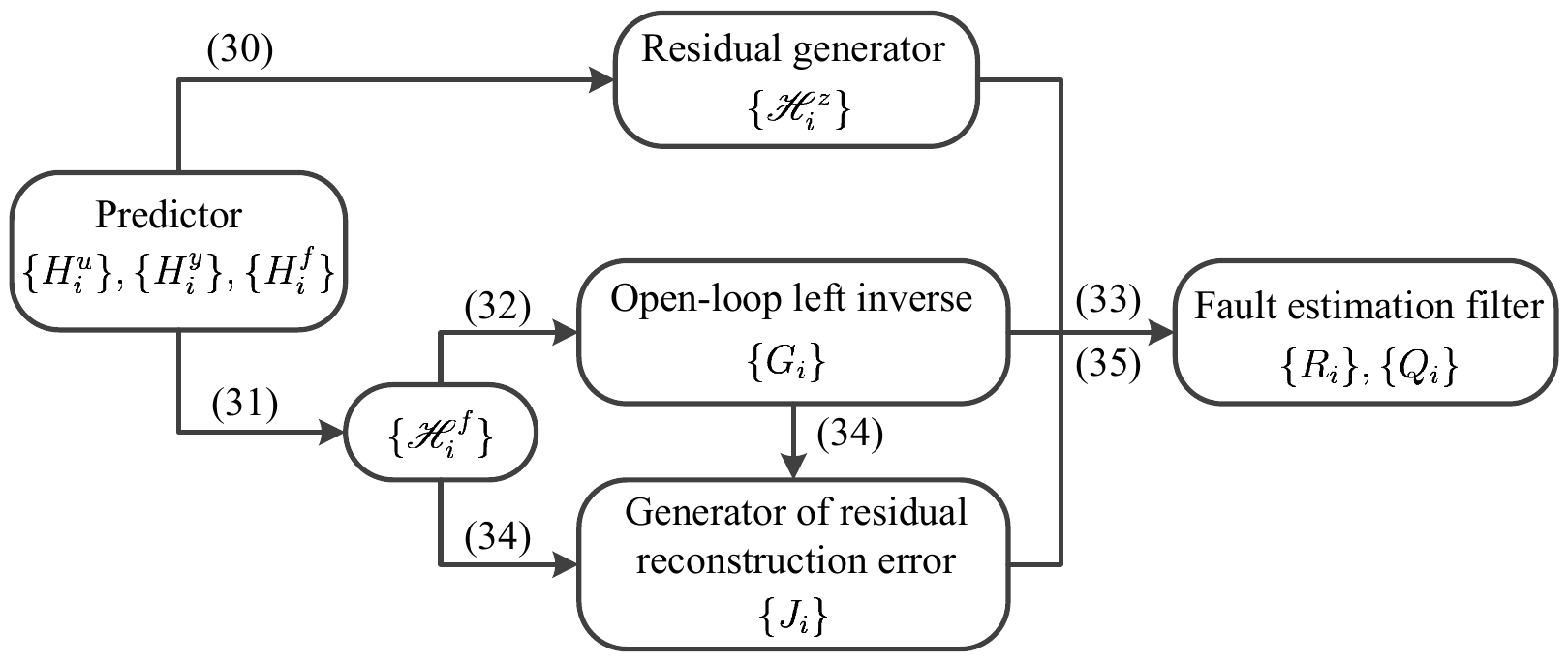}    
		\caption{Link between predictor MPs and MPs of SI-FEF}
		\label{fig:MPlink}
	\end{center}
\end{figure}

Equations (\ref{eq:scrHz})-(\ref{eq:Qi}) reveal the link from the predictor MPs to the MPs of the SI-FEF (\ref{eq:sfestfilter_reduced}), as summarized in Figure \ref{fig:MPlink}.

\section{Fault estimation filter design using Markov parameters}\label{sect:design_using_Markov}
By exploiting the link between the predictor MPs and the SI-FEF MPs, as analyzed in Section \ref{sect:fefextend}, the proposed MP based data-driven design is given as below.

\begin{description}[font=\normalfont\textit]
	\item[Algorithm A. Data-driven design of fault estimation filter] 
\end{description}
\begin{enumerate}[label=(\roman*)]
	\item Identify the predictor MPs $\{ H_i^u\}$ and $\{ H_i^y \}$ using VARX modelling with the historical or experimental fault-free I/O data.
	\item Compute MPs of SI-FEF (\ref{eq:sfestfilter_reduced}). \\
	Construct the MPs $\{H_i^f\}$, $\{\mathscr{H}_i^z\}$, and $\{\mathscr{H}_i^f\}$ according to (\ref{eq:Hif}), (\ref{eq:scrHz}), and (\ref{eq:scrHf}), respectively. Select one left inverse matrix $\Pi$ of $H_\tau^f$, e.g., $\Pi = \left( (H_\tau^f)^\text{T} H_\tau^f \right)^{-1} (H_\tau^f)^\text{T}$. Then compute $\{G_i\}$, $\{J_i\}$, $\{R_i\}$, and $\{Q_i\}$ by following (\ref{eq:Gi})-(\ref{eq:Qi}). 
	\item State-space realization of the SI-FEF (\ref{eq:sfestfilter_reduced}) from the MPs $\{R_i\}$ and $\{Q_i\}$. \\
	According to (\ref{eq:RL_QL}) and (\ref{eq:RL_QL_Markov}), the MPs $\{R_i\}$ and $\{Q_i\}$ correspond to systems 
	$\left( \Phi_1, [\begin{matrix} B_f & K_f \end{matrix}], C_1,
	[\begin{matrix} D_{f,1} & G_{f,1} \end{matrix}] \right)$ and 
	$\left( \Phi_1, [\begin{matrix} B_f & K_f \end{matrix}], -C_2,
	[\begin{matrix} -D_{f,2} & G_{f,2} \end{matrix}] \right)$, respectively. Then it is straightforward to obtain
	\begin{equation*}
	\left[ \begin{matrix} \hat D_{f,1} & \hat G_{f,1} \end{matrix} \right] = R_0, \,
	\left[ \begin{matrix} - \hat D_{f,2} & \hat G_{f,2} \end{matrix} \right] = Q_0.
	\end{equation*}
	Formulate two block-Hankel matrices $\mathcal{H}_{R}$ and $\mathcal{H}_{Q}$ as
	\begin{equation}\label{eq:HRL}
	\mathcal{H}_{W} = \left[
	\begin{smallmatrix}
	W_1 & W_2 & \cdots & W_m \\
	W_2 & W_3 & \cdots & W_{m+1} \\
	\vdots & \vdots & \ddots & \vdots \\
	W_l & W_{l+1} & \cdots & W_{l+m-1}
	\end{smallmatrix}
	\right], \, W \text{ represents } R \text{ or } Q,
	\end{equation}
	then compute their singular value decomposition (SVD), i.e., 
	\begin{equation*}\label{eq:HRsvd}
	\mathcal{H}_{W} = \left[ \begin{matrix} U_W & U_W^{\bot} \end{matrix} \right]
	\left[ \begin{matrix}
	\Sigma_W & 0 \\
	0 & \Sigma_W^{\bot}
	\end{matrix} \right]
	\left[ \begin{matrix} V_W^\mathrm{T} \\ \left( V_W^{\bot} \right)^\mathrm{T} \end{matrix} \right].
	\end{equation*}	
	In this above equation, the nonsingular and diagonal matrices $\Sigma_R$ and $\Sigma_Q$ consist of the $\hat n$ largest singular values of $\mathcal{H}_R$ and $\mathcal{H}_Q$, respectively, where $\hat n$ is the selected order of the fault estimation filter (\ref{eq:sfestfilter_reduced}). The order $\hat n$ can be chosen by examining the gap among the singular values of $\mathcal{H}_R$ and $\mathcal{H}_Q$, respectively, as in subspace identification methods \cite{Chiuso2007}. Let the rank-reduced block-Hankel matrices $\mathcal{\hat H}_R$ and $\mathcal{\hat H}_Q$ be
	\begin{equation}\label{eq:HR_HQ_app}
	\mathcal{\hat H}_{W} = U_W \Sigma_W V_W^\mathrm{T},
	\, W \text{ represents } R \text{ or } Q.
	\end{equation}
	
	For $\mathcal{\hat H}_R$ defined in (\ref{eq:HR_HQ_app}), the estimated controllability and observability matrices can be constructed as \cite{Chiuso2007}
	\begin{equation}\label{eq:Ctrb_Obsv}
	\mathcal{\hat C}_R = \Sigma_R^{\frac {1}{2}} V_R^\mathrm{T}, \;
	\mathcal{\hat O}_R = U_R \Sigma_R^{\frac {1}{2}}. \;
	\end{equation}
	Then the state-space realization of $\mathcal{\hat H}_R$ are computed as below:
	\begin{align}
	[\begin{matrix} \hat B_f & \hat K_f \end{matrix}] &= \text{the first}\;{n_u+n_y}\;\text{columns of}\;\mathcal{\hat C}_R, \nonumber \\
	\hat C_1 &=
	\text{the first}\;{n_f}\;\text{rows of}\;\mathcal{\hat O}_R, \nonumber \\
	\hat \Phi_1 &= \mathcal{\hat C}_{R,2} \mathcal{\hat C}_{R,1}^\mathrm{T} \left( \mathcal{\hat C}_{R,1} \mathcal{\hat C}_{R,1}^\mathrm{T} \right)^{-1}, \nonumber
	\end{align}
	where $\mathcal{\hat C}_{R,1}$ and $\mathcal{\hat C}_{R,2}$ are the matrices consisting of the first and, respectively, the last $n_u \left( m-1 \right)$ columns of $\mathcal{\hat C}_{R}$.
	According to (\ref{eq:RL_QL}), the state-space realizations of the block-Hankel matrices $\mathcal{\hat H}_R$ and $\mathcal{\hat H}_Q$ have the same controllability matrix, i.e., $\mathcal{\hat C}_R$ obtained in (\ref{eq:Ctrb_Obsv}). 
	Then the observability matrix in the state-space realization of $\mathcal{\hat H}_Q$ can be computed below by using $\mathcal{\hat H}_Q = \mathcal{\hat O}_Q \mathcal{\hat C}_R$:
	\begin{equation}\label{eq:OQ}
		\mathcal{\hat O}_Q = \mathcal{\hat H}_Q \mathcal{\hat C}_R^\text{T} \left( \mathcal{\hat C}_R \mathcal{\hat C}_R^\text{T} \right)^{-1}.
	\end{equation}
	Finally, we have 
	\begin{equation}\label{eq:C2_hat}
		- \hat C_2 =
		\text{the first}\;{n_y}\;\text{rows of}\;\mathcal{\hat O}_Q. 
	\end{equation}
%
	\item Design the filter gain $K_r$ by following Algorithm B in Section \ref{sect:gain_design}; and construct the SI-FEF (\ref{eq:sfestfilter_reduced}) with the identified system matrices in Step (\romannumeral3). 
\end{enumerate}

\begin{rem}\label{rem:L_l_m}
The VARX model order $p$ in Step (\romannumeral1) is selected according to Remark \ref{rmk:p}.
In Step (\romannumeral2), the length $L$ of the SI-FEF MPs needs to be sufficiently large to ensure satisfactory fault estimation performance. This is due to the asymptotic unbiasedness of the batch fault estimation (\ref{eq:festfilter_extended0}) as $L$ goes to infinity, which is explained in Section \ref{sect:fefextend}.
In Step (\romannumeral3), we select the size of the block-Hankel matrix in (\ref{eq:HRL}) to be $l+m = L$, with $l$ and $m$ defined in (\ref{eq:HRL}). By doing so, all MPs $\{R_i, Q_i\}$ ($i=1, 2, \cdots, L$) obtained in Step (\romannumeral2) are used to construct $\mathcal{H}_R$ and $\mathcal{H}_Q$ in (\ref{eq:HRL}).
\end{rem}

\subsection{Suboptimal design of filter gain}\label{sect:gain_design}
The joint design of both $\Pi$ in (\ref{eq:Pi_matrix}) and the filter gain $K_r$ is extremely difficult, because all system matrices in the SI-FEF (\ref{eq:sfestfilter_reduced}) depend on $\Pi$. Alternatively, our proposed data-driven design selects
$\Pi$ in Step (\romannumeral2) of Algorithm A before designing the steady-state filter gain $K_r$ in Step (\romannumeral4) of Algorithm A. Therefore, designing the filter gain $K_r$ given a predefined $\Pi$ is suboptimal compared to the joint design. Recall that in some existing model-based unknown input estimation methods,  $\Pi$  and $K_r$ were jointly designed to achieve globally unbiased minimum-variance estimation, e.g., in \cite{Gill2007}, without discussing stability of the obtained filter therein.

Based on the fault estimation error dynamics (\ref{eq:fest_err_dyn}), the $\mathcal{H}_2$ fault estimation problem can be formulated as
\begin{equation}\label{eq:H2opt1}
\min\limits_{{K}_r} \| \hat{C}_1 (zI - \hat{\Phi}_1 + {K}_r \hat{C}_2)^{-1} (\hat{B}_1 + {K}_r \hat{D}_2) \Sigma_e^{\frac{1}{2}} \|_2^2
\end{equation}
to find the steady-state filter gain ${K}_r$.
It is well known that the solution ${K}_r$ to the problem (\ref{eq:H2opt1}) does not depend on $\hat{C}_1$, and is actually the steady-state Kalman filter gain, see Section 6.5 of \cite{Kwad1972} and Section 7.3 of \cite{Burl1998}.
In this above problem formulation,
$\hat \Phi_1$, $\hat C_1$, and $\hat C_2$ are obtained in Algorithm A as the estimates of $\Phi_1$, $C_1$, and $C_2$, respectively, while estimating $\hat B_1$ and $\hat D_2$ will be explained later in Step (\romannumeral1) of Algorithm B. 

With these above estimated matrices, the solution to the problem (\ref{eq:H2opt1}) is discussed as below. 
Note that in Step (\romannumeral1) of Algorithm B, we have
\begin{equation}\label{eq:breveD2}
\hat{D}_2 = J_0 = I - {H}_{\tau}^f \Pi
\end{equation}
according to (\ref{eq:scrHf}), (\ref{eq:Gi}), and (\ref{eq:Ji}), and we have $\Pi \hat{D}_2 = 0$ since $\Pi {H}_{\tau}^f = I$. Then it can be seen that $\hat{D}_2$ is row-rank deficient, hence the solution to (\ref{eq:H2opt1}) is non-unique. To tackle this problem, we follow \cite{Darou1997} to restrict the filter gain ${K}_r$ to be in the form 
\begin{equation*}
{K}_r = \bar K_r \alpha,
\end{equation*} 
where $\alpha \in \mathbb{R}^{s \times n_y}$ ensures $\text{rank}(\hat{D}_2) = \text{rank}(\alpha \hat{D}_2) = s$. Then the $\mathcal{H}_2$ optimization problem (\ref{eq:H2opt1}) becomes
\begin{equation}\label{eq:H2opt2}
\min\limits_{\bar K_r} \| \hat{C}_1 (zI - \hat{\Phi}_1 + \bar K_r \bar C_2)^{-1} (\hat{B}_1 + \bar K_r \bar D_2) \Sigma_e^{\frac{1}{2}} \|_2^2
\end{equation}
with $\bar C_2 = \alpha \hat{C}_2$ and $\bar D_2 = \alpha \hat{D}_2$. 
With a proper selection of $\alpha$, the sufficient and necessary condition given below in Theorem \ref{thm:ARE} guarantees that the solution to (\ref{eq:H2opt2}), i.e., 
\cite{Kwad1972,Burl1998}
\begin{equation}\label{eq:barKr}
\bar K_r = \left( \hat{\Phi}_1 P \bar C_2^\text{T} + \hat{B}_1 \Sigma_e \bar D_2^\text{T} \right) 
\Xi_e^{-1}
\end{equation}
stabilizes the SI-FEF (\ref{eq:sfestfilter_reduced}), where $P$ is the stabilizing solution to the algebraic Riccati equation (ARE)
\begin{subequations}\label{eq:ARE}
	\begin{align}
	P &= \hat{\Phi}_1 P \hat{\Phi}_1^\text{T} + \hat{B}_1 \Sigma_e \hat{B}_1^\text{T} \\
	&\quad- \left( \hat{\Phi}_1 P \bar C_2^\text{T} + \hat{B}_1 \Sigma_e \bar D_2^\text{T} \right) 
	\Xi_e^{-1}
	\left( \hat{\Phi}_1 P \bar C_2^\text{T} + \hat{B}_1 \Sigma_e \bar D_2^\text{T} \right)^\text{T}, \nonumber\\
	\Xi_e &= \bar C_2 P \bar C_2^{\text{T}} + \bar D_2 \Sigma_e \bar D_2^{\text{T}}.
	\end{align}
\end{subequations}

\begin{lem}\label{lem:alpha}
The selected $\alpha$ in Step (\romannumeral2) of Algorithm B ensures that (\romannumeral1) the matrix $\left[ \begin{matrix}
\alpha \\ \Pi
\end{matrix} \right]$ is nonsingular; and (\romannumeral2) $\Pi \hat {C}_2 = 0$.
\end{lem}
The proof is given in the Appendix.  \qed

Despite the identification errors in $\hat C_2$ compared to the true $C_2$ defined in (\ref{eq:C2_D2}), the condition (\romannumeral2) of Lemma \ref{lem:alpha} still holds, just like the fact that $\Pi  {C}_2 = 0$ holds for the true $C_2$. This condition will be used in the proof of Theorem \ref{thm:ARE} below.

\begin{theorem}\label{thm:ARE}
With Assumption \ref{ass:fault_rank} and the selection of $\alpha$ in Step (\romannumeral2) of Algorithm B, the ARE (\ref{eq:ARE}) has a unique stabilizing solution $P$ if and only if 
\begin{subequations}
\begin{align}
& \text{rank} \left[ \begin{matrix}
\hat \Phi_1 - \lambda I \\
\hat C_2
\end{matrix} \right] = n, \text{ for } |\lambda| \geq 1, \label{eq:obsvb} \\
& \text{rank} \left[ \begin{matrix}
\hat{\Phi}_1 - e^{j\omega} I & \hat{B}_1 \\
\hat{C}_2 & \hat{D}_2
\end{matrix} \right] = n + n_y, \text{ for } \omega \in [0,2\pi]. \label{eq:contrl}
\end{align}
\end{subequations}
\end{theorem}
The proof is given in the Appendix.  \qed

Theorem \ref{thm:ARE} shows that the existence of a unique stabilizing solution to the $\mathcal{H}_2$ estimation problem (\ref{eq:H2opt2}) depends on the system matrices of $(\hat \Phi_1, \hat B_1, \hat C_2, \hat D_2)$, despite errors in these identified matrices. 
Such a suboptimal design and its stability guarantee are not provided in recently proposed data-driven fault estimation observers in \cite{Dong2012c} and Chapter 10 of \cite{Ding2014book}.


The design method for the filter gain $K_r$
is now summarized in Algorithm B.

\begin{description}[font=\normalfont\textit]
	\item[Algorithm B. Suboptimal design of filter gain] 
\end{description}
\begin{enumerate}[label=(\roman*)]
	\item Identify $B_1$ and $D_2$ using the MPs $\{J_i\}$ identified in the Step (\romannumeral2) of Algorithm A. \\
		From (\ref{eq:JL}) and (\ref{eq:RL_QL_Markov}), we can see that $\{J_i\}$ are the MPs of the system $\left( \Phi_1, B_1, -C_2, D_2 \right)$.
		It is then easy to obtain $\hat D_2 = J_0$. Formulate the block-Hankel matrix $\mathcal{H}_J$ with the MPs $\{J_i\}$ by using the definition (\ref{eq:HRL}). With the selected filter order $\hat n$, we compute the rank-reduced matrix $\mathcal{\hat H}_J$ by following procedures similar to Step (\romannumeral3) of Algorithm A. Since the observability matrix of the state-space realization of $\mathcal{\hat H}_J$ is the same as that of $\mathcal{\hat H}_Q$, i.e.,
		$\mathcal{\hat O}_Q$ in (\ref{eq:OQ}), we can compute the controllability matrix $\mathcal{\hat C}_J$ of $\mathcal{\hat H}_J$ as below by using $\mathcal{\hat H}_J = \mathcal{\hat O}_Q \mathcal{\hat C}_J$: 
		\begin{equation*}
			\mathcal{\hat C}_J = ( \mathcal{\hat O}_Q^\text{T} \mathcal{\hat O}_Q  )^{-1} \mathcal{\hat O}_Q^\text{T} \mathcal{\hat H}_J.
		\end{equation*}
		Finally, we obtain $\hat B_1$ as the first $n_y$ columns of $\mathcal{\hat C}_J$.
	\item Let the SVD of $H_\tau^f$ be 
	\begin{equation*}
	H_\tau^f = \left[\begin{matrix}
	U_1 & U_2
	\end{matrix}\right] \left[\begin{matrix}
	S_H \\ 0
	\end{matrix}\right] V^\text{T},
	\end{equation*}
	then we select $\alpha = U_2^\text{T}$ so that $\alpha \hat{D}_2 = \alpha (I - {H}_{\tau}^f \Pi) = U_2^\text{T}$ is full row rank according to (\ref{eq:breveD2}).
	\item With $\bar C_2 = \alpha \hat C_2$ and $\bar D_2 = \alpha \hat D_2$, compute $\bar K_r$ in (\ref{eq:barKr}) by solving the ARE (\ref{eq:ARE}). Then the filter gain is ${K}_r = \bar K_r \alpha$. 
\end{enumerate}

\subsection{Comparisons and discussions}\label{Sect:comp_disc}
The data-driven FIR fault estimator have been reviewed in Section \ref{sect:prob_form_C}. Next, we focus on further comparisons with other existing state-space FEF designs from data.

The data-driven method of \cite{Dong2012c} considered only the open-loop left inverse $\mathcal{G}_L$ (\ref{eq:GL}) corresponding to (\ref{eqr:leftinv_1}) with $K_r = 0$. Hence, it has no stability guarantees. In contrast, our data-driven design is based on the closed-loop left inverse (\ref{eqr:leftinv_1}) and its extended form (\ref{eq:festfilter_extended0}) which ensure the stability and asymptotical unbiasedness under the condition specified in Theorem \ref{thm:stabilizability} and explained in Section \ref{sect:fefextend}.

For the above reason, the data-driven filter in \cite{Dong2012c} cannot be applied to sensor faults of an unstable open-loop plant.
It is worth noting that this difficulty cannot be solved by simply applying the same method to the stabilized closed-loop system. The reason is that the sensor faults affect not only the output equations but also the closed-loop dynamics, hence (\ref{eq:Hif}) is no longer valid for the MPs $\{ H_i^f \}$ of the closed-loop fault subsystem. Section 2 of \cite{Wan2012} provides more detailed analysis about the effect of sensor fault propagation in closed-loop dynamics. To circumvent this difficulty, Section \uppercase\expandafter{\romannumeral5}-B of \cite{Dong2012c} proposed to use a special control law such that the sensor faults did not affect the closed-loop dynamics, which is not always possible in practice.

With the parity vector identified from data, Chapter 10 of \cite{Ding2014book} constructed a diagnostic observer and estimated faults as augmented state variables. This augmented observer scheme, however, imposed certain limitations on how fault signals vary with time, thus introduced bias in fault estimates. In contrast, our proposed data-driven design needs no assumptions about the time-varying fault signals. Moreover, for systems with multiple outputs, the parity vector based approach in Chapter 10 of \cite{Ding2014book} becomes much complicated, while our proposed method remains the same.

The suboptimality of our data-driven design is mainly due to three reasons: firstly, the identification errors of the predictor MPs in Step (\romannumeral1) of Algorithm A is neglected, as pointed out in the last paragraph of Section \ref{sect:prob};
secondly, the unmodelled dynamics of the state-space realization in Step (\romannumeral3) of Algorithm A is not explicitly considered; 
and thirdly, the matrix $\Pi$ is not jointly designed with the filter gain, as explained in the first paragraph of Section \ref{sect:gain_design}. 

Further comments are in order for the second reason above in our suboptimal design. 
The unmodelled dynamics in our proposed approach is the result of approximating the batch fault estimator (\ref{eq:festfilter_extended}) with a state-space filter.
The higher state order of the designed filter leads to better approximation, thus giving better fault estimation performance.
Because of this, the obtained state-space filter cannot achieve better performance than the batch estimator (\ref{eq:festfilter_extended}) given a fixed horizon length $L$.
Therefore, considering also heavier online computational load due to a higher state order, the order determination of our designed filter is a simple trade-off between the fault estimation performance and the computational load. 
In contrast, the order determination is cumbersome for the conventional two-step design and the parity vector based design in Chapter 10 of \cite{Ding2014book}. 
In these two approaches, we need to select the order of a state-space plant model or a parity vector that represents a residual subspace. The cumbersome issue is that the model mismatch is introduced in the very first step of these two approaches, and propagates through all the sequential steps. Due to this complicated error propagation, there are no clear guidelines for selecting the state order for the fault estimation performance. 

The two sources of uncertainties described in the first two reasons of the suboptimality of our proposed approach are common in most existing data-driven design methods. How to explicitly quantify and deal with their effects remains an open problem. One solution to the identification errors of MPs has been investigated in \cite{WanKevicVerh2016} by using a nonrecursive receding horizon estimator. However, it is still a challenge to address both sources of uncertainties mentioned above in the data-driven design of a recursive FEF filter.

\section{Simulation studies}\label{sect:sim}
Consider the linearized continuous-time vertical takeoff and landing aircraft model used in \cite{Dong2012c}.
The model has four states, namely horizontal velocity, vertical velocity, pitch rate, and pitch angle. The two inputs are collective pitch control and longitudinal cyclic pitch control, 
both of which are fed through the second-order actuator 
$$
\frac{21.3501s+162.3867}{s^2 + 17.9994s + 162.3867}.
$$
The sampling interval is 0.5 second.
The process and measurement noises are zero mean white, with covariances of $10^{-4} I_4$ and $0.0016 I_2$ respectively.
Since the open-loop plant is unstable, an empirical stabilizing output feedback controller is used, i.e.,
\begin{equation}\label{eq:controller}
u(k) = - \left[ \begin{matrix}
0 & 0 & -0.5 & 0 \\
0 & 0 & -0.1 & -0.1
\end{matrix}
\right] y(k) + \eta(k),
\end{equation}
where $\eta(k)$ is the reference signal. All the parameters of the plant and the controller are the same as those in \cite{Dong2012c}.
The plant model is unknown to our data-driven design problem.

In the identification experiment, the reference signal $\eta(k)$ is zero-mean white noise with the covariance of $\mathrm{diag}\left( 1, 1 \right)$, which ensures persistent excitation.
We collect $N=100000$ data samples, and select the VARX model order to be $p=12$ by following Remark \ref{rmk:p}.

The simulated fault scenarios include: 1) faults in the two actuators; 2) faults in the first two sensors.
In both scenarios, the reference signal $\eta (k)$ in the control law (\ref{eq:controller}) is set to be $\left[\begin{matrix} 2 & 2 \end{matrix}\right]^\mathrm{T}$, and the fault signals are 
$$ f(k) = \left\{ \begin{array}{ll}
\left[ \begin{array}{cc}
0 & 0
\end{array}
\right]^\mathrm{T}, & 0 \leq k \leq 500, \\
\left[ \begin{array}{cc}
1 & \mathrm{sin}\left( 0.01 \pi k \right) 
\end{array}
\right]^\mathrm{T}, & k > 500.
\end{array}
\right. $$

We will compare the following methods for data-driven FEF design:
\begin{itemize}
	\item Alg0: the SI-FEF (\ref{eq:sfestfilter_reduced}) using the accurate predictor model (\ref{eq:predictor});
	\item Alg1: the SI-FEF (\ref{eq:sfestfilter_reduced}) using the state-space model of the predictor (\ref{eq:predictor}) identified from data;
	\item Alg2: the data-driven method proposed in \cite{Dong2012c};
	\item Alg3: our proposed new method in Section \ref{sect:design_using_Markov}.
\end{itemize}
The first algorithm is model-based, and the other three methods are all data-driven.
For Alg2 and Alg3, the length of the time window to construct the data-driven FEF is $L = 90$, and the number of block rows and columns of the block-Hankel matrix $\mathcal{H}_W$ in (\ref{eq:HRL}) is $l = m = 45$, according to the guidelines in Remark \ref{rem:L_l_m}.

\begin{figure}[!h]
	\centering
	\subfigure[State order: 8]{
		\includegraphics[width=8cm]{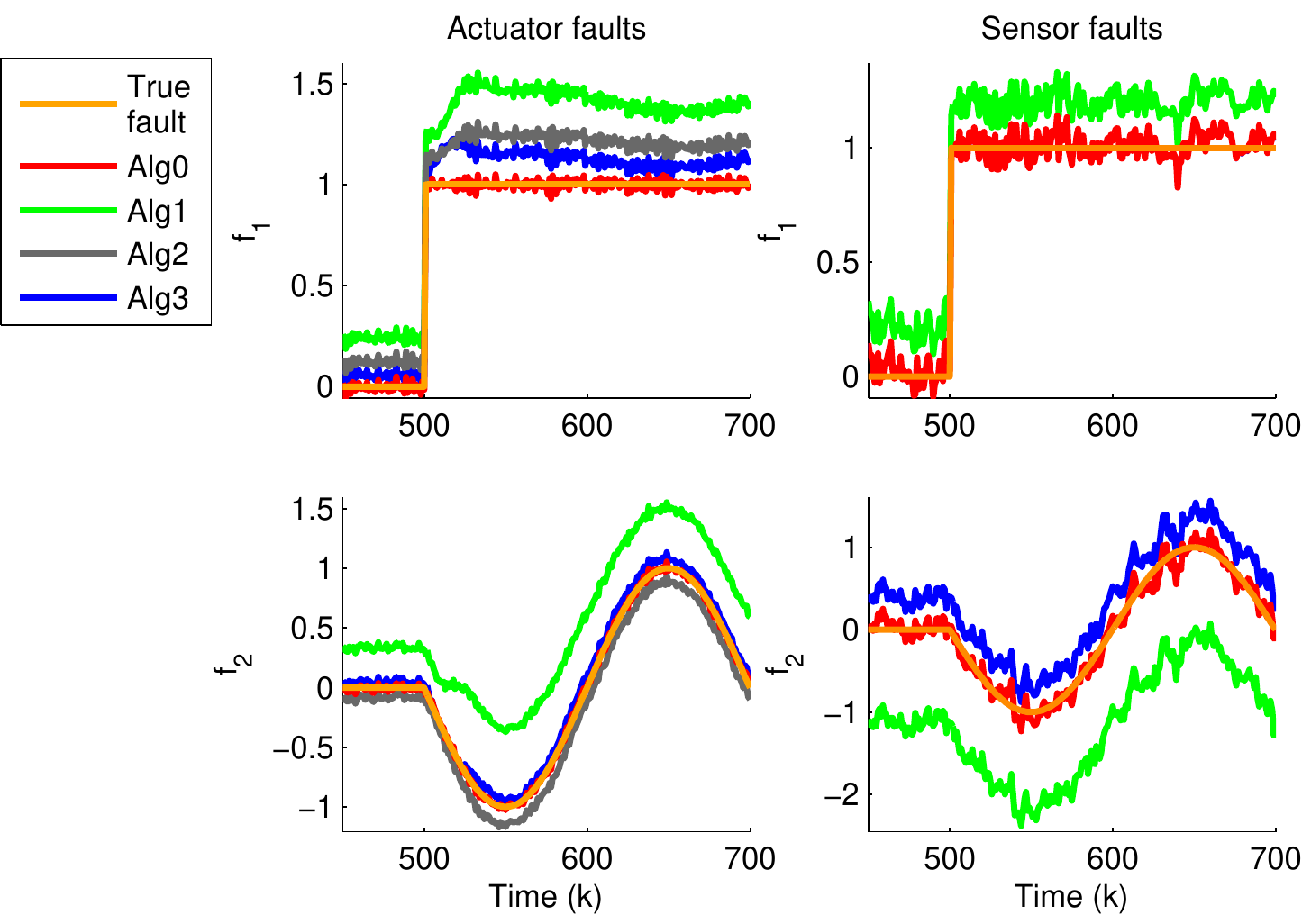}
	}
	\subfigure[State order: 18]{
		\includegraphics[width=8cm]{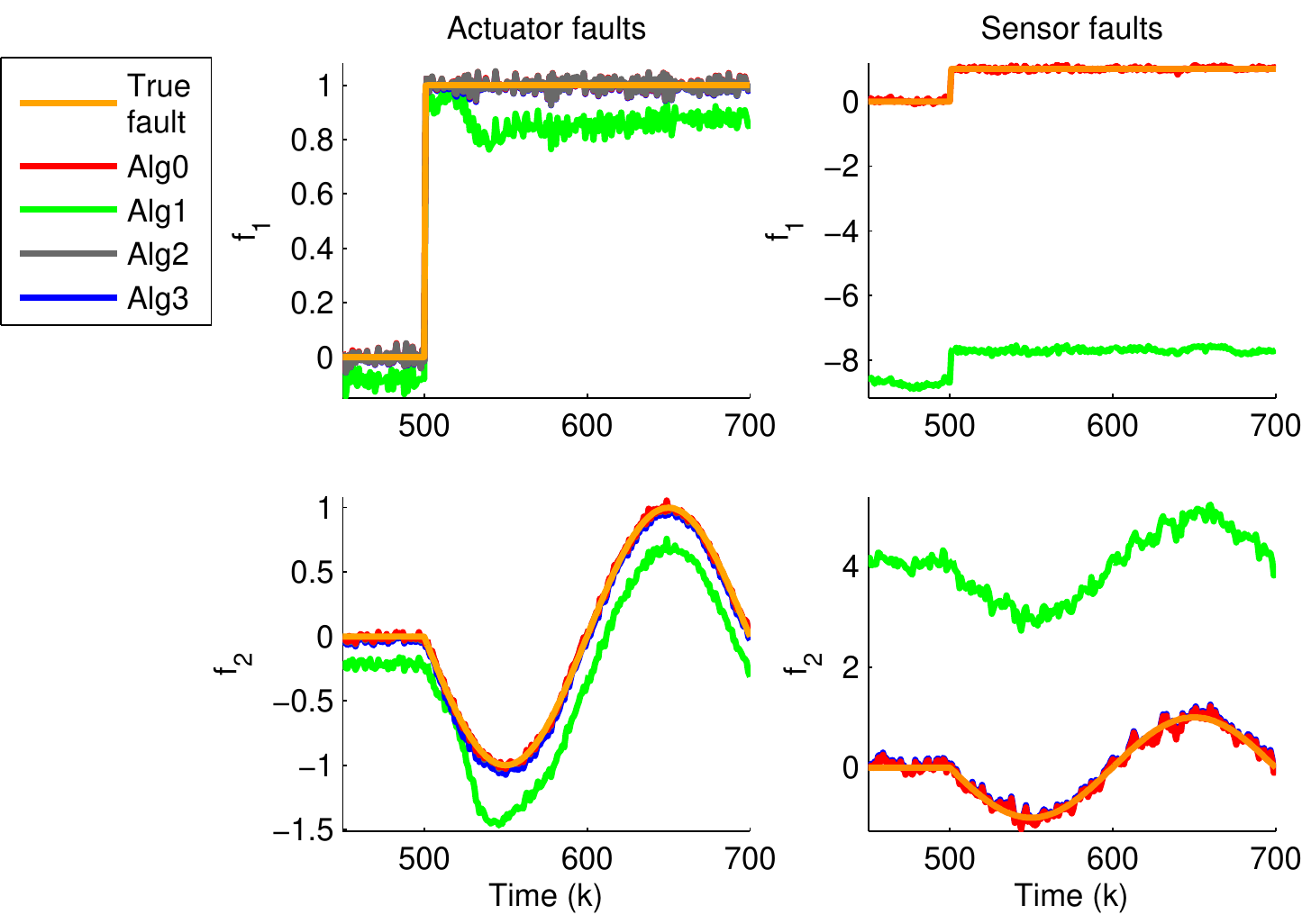}
	}
	\caption{Fault estimates given by different methods. (The result of Alg2 is not plotted for sensor faults because it gives an unstable filter.)}
	\label{fig:fest}
\end{figure}

First, all state orders in the three data-driven designs are set to be 8, i.e., the true state order of the underlying system. 
By doing so, we focus on the stability of the obtained filters.
The estimated fault signals are shown in Figure \ref{fig:fest}(a). 
With the accurate plant model, we can see that the fault subsystem has stable invariant zeros in the actuator fault scenario, and no invariant zeros in the sensor fault scenario. 
Therefore, the model-based approach Alg0 gives stable FEFs in both faulty scenarios according to Theorems \ref{thm:stabilizability} and \ref{thm:ARE}.
However, Alg2 results in an unstable FEF in the sensor fault scenario due to the reason explained in Section \ref{Sect:comp_disc}, thus it is not plotted for the sensor faults. In contrast to Alg2, Alg1 and Alg3 are based on the closed-loop left inverse (\ref{eqr:leftinv_1}), and their stability is guaranteed under the conditions in Theorems \ref{thm:stabilizability} and \ref{thm:ARE}. 
Because of the model identification errors multiplying with the online I/O data, the three data-driven designs, i.e., Alg1, Alg2, and Alg3, all give larger estimation errors than Alg0. In comparison, 
Alg2 and Alg3 suffer less from model identification errors than Alg1.

\begin{figure}[h]
	\centering
	\includegraphics[width=9cm]{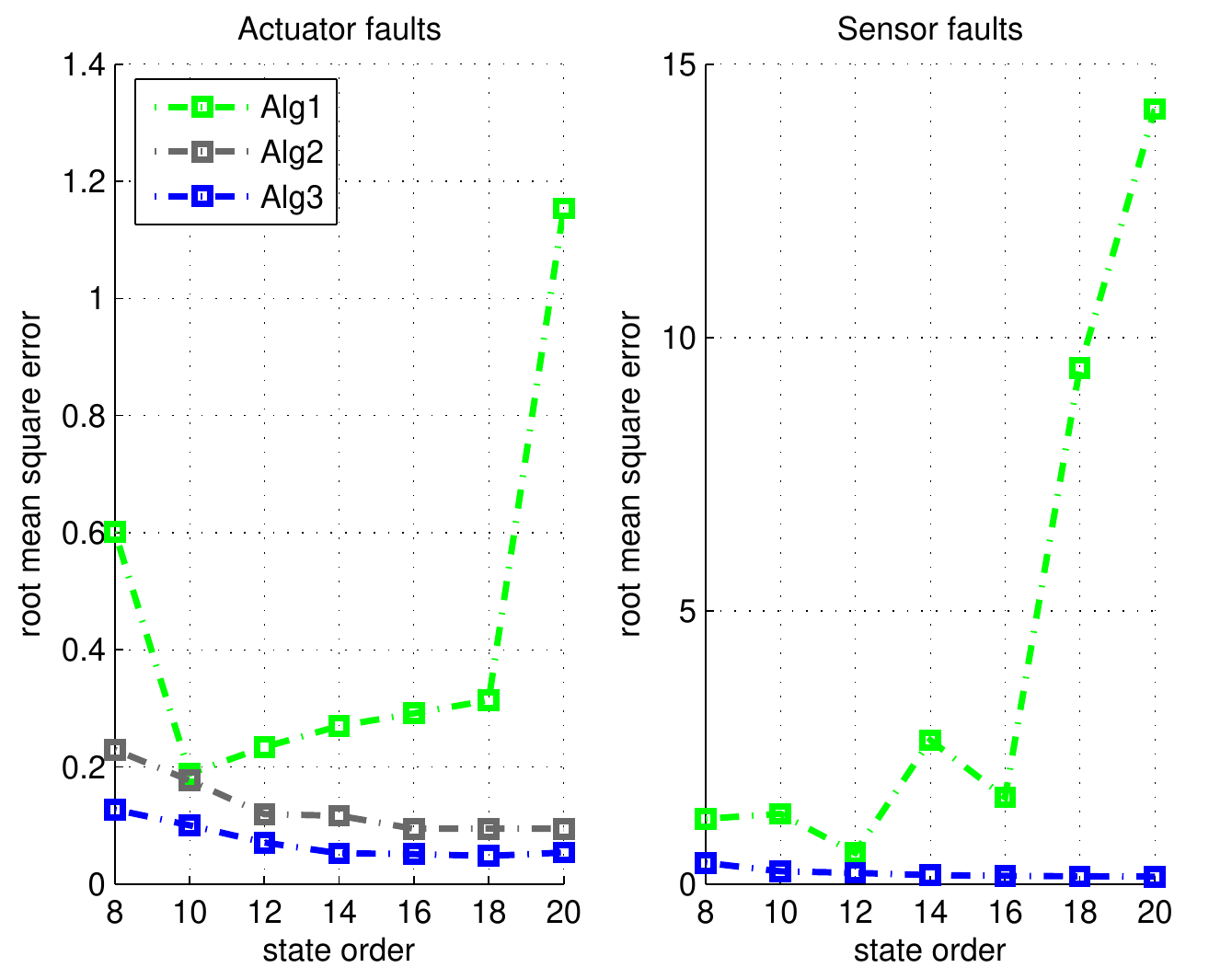}
	\caption{Root mean square error of fault estimates when selecting different state orders. (The results of Alg2 are not plotted for sensor faults because it gives unstable filters.)}
	\label{fig:order}
\end{figure}

Next, we examine the state order selection in the data-driven designs.
For Alg1, the state order is determined in the phase of state-space plant model identification, which results in unavoidable model mismatch. 
Due to complicated error propagation from model mismatch, the fault estimation performance may drastically change with different state orders, as illustrated in Figures \ref{fig:fest}(b) and \ref{fig:order} when different state orders from 8 to 20 are used. Moreover, Figure \ref{fig:order} shows even the selection of the true plant state order 8 may not necessarily give smaller fault estimation errors. 
In contrast, the state-space filters in Alg2 and Alg3 are both approximations to batch fault estimators.
Then the selection of a higher state order in Alg2 and Alg3 leads to better approximation, thus gives better fault estimation performance in Figures \ref{fig:fest}(b) and \ref{fig:order}.
This comparison shows that it is much easier to determine the state order of our proposed design: choose a high state order as long as the online computational load is allowed. 
More detailed reasons are explained in the second-to-last paragraph of Section \ref{Sect:comp_disc}.

\begin{figure}[h]
	\centering
	\includegraphics[width=9cm]{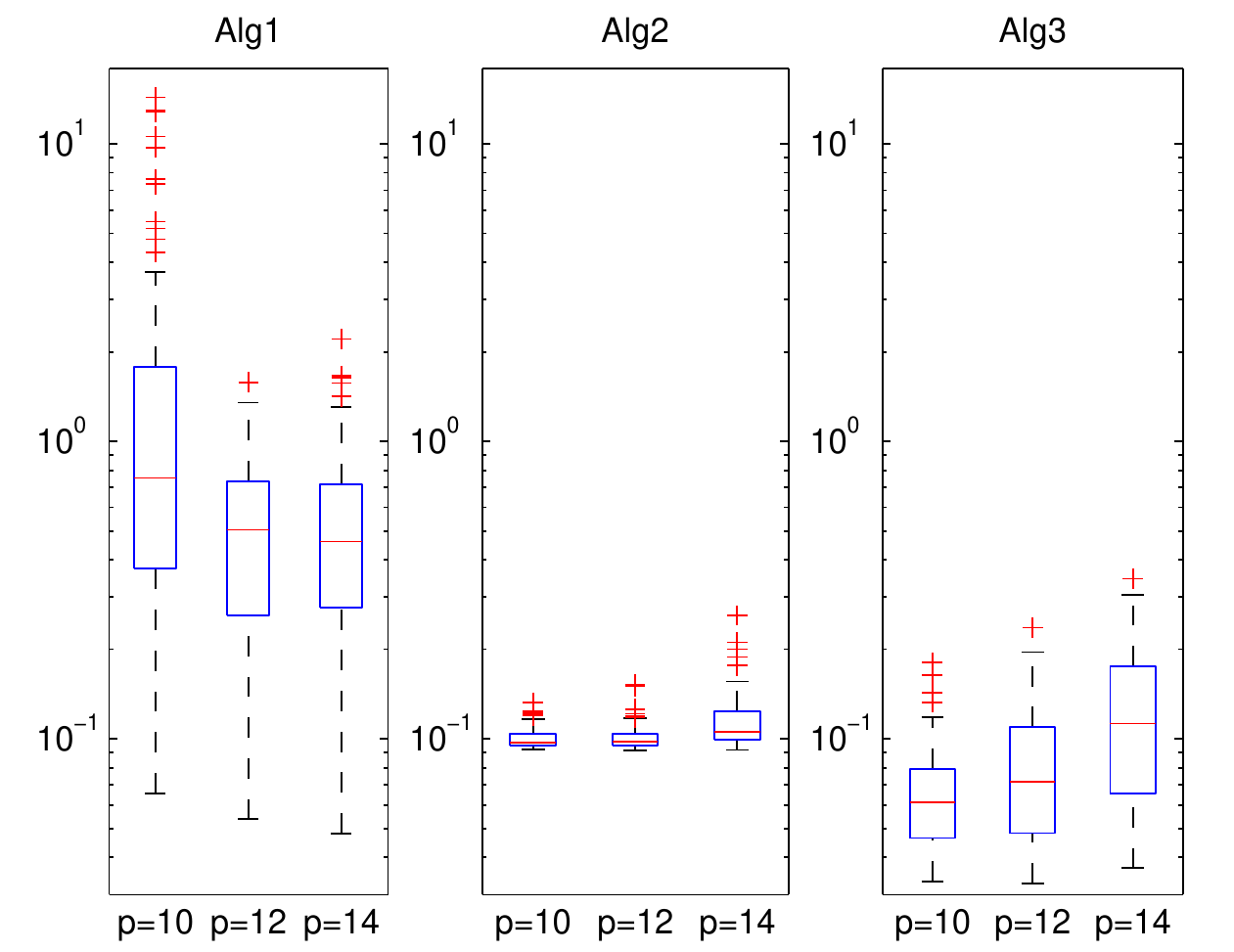}
	\caption{Boxplots of root mean square error of fault estimates in Monte Carlo simulations of the actuator fault scenario: 100 Monte Carlo runs for each different selection of the VARX order $p$; the state orders of Alg1, Alg2, and Alg3 are set to 8, 14, and 14, respectively.}
	\label{fig:MC_p}
\end{figure}

Besides the state order selection discussed above, the VARX order $p$ is also critical, since it is related to the accuracy of the identified predictor MPs. For different VARX order selections $p=10, 12, 14$, we implement 100  Monte Carlo runs for each $p$, with the state orders of Alg1, Alg2, and Alg3 set to 8, 14, and 14, respectively. The results are shown as boxplots in Figure \ref{fig:MC_p}.
With the VARX order increasing, the identified predictor MPs have smaller biases but larger variances. 
This results in slightly worse performance for Alg2 and Alg3 when a larger VARX order is used. 
In comparison, our proposed Alg3 gives the smallest averaged root mean square error of fault estimates, Alg2 has a larger averaged root mean square error but the smallest variance, and Alg1 gives the worst performance. 
In the above Monte Carlo runs of the actuator fault scenario, our proposed Alg3 does not consistently perform better than Alg2 for all selections of the VARX order $p$. However, it should be noted that in all the Monte Carlo runs of the sensor fault scenario, Alg3 can stabilize the designed filter whereas Alg2 fails to do so.

\section{Conclusions}\label{sect:conclusion}
A novel direct data-driven design method has been proposed for FEFs by parameterizing the system-inversion-based fault estimation filter with predictor Markov parameters.
The proposed approach does not need to identify a state-space plant model, but still allows the filter gain design for stabilization and suboptimal $\mathcal{H}_2$ performance. This has not been achieved by other existing data-driven fault estimation methods so far. Moreover, the fault estimation performance can be improved by simply increasing the state order of the designed filter, at the cost of higher online computational load.
A numerical simulation example illustrates the advantages of our method applied to actuator and sensor faults of an unstable aircraft system.
Future work will focus on the robustification of our data-driven design against identification errors of Markov parameters, and non-Gaussian distributions in real applications.

\appendix
\section*{Proof of Theorem \ref{thm:stabilizability}}\label{app:proof_stabilizability}    
In order to prove $( \Phi_1, C_2 )$ is detectable, we need to show that $( \Phi_1, C_2 )$ has no unstable unobservable modes, i.e.,
\begin{equation}\label{eq:unobsv_mode}
\text{rank} \left[ \begin{array}{c}
\Phi_1 - \lambda I \\
C_2
\end{array}
\right] = n \;\text{for}\; \left| \lambda \right| \ge 1.
\end{equation}

By following (\ref{eq:Phi1_B1}) and (\ref{eq:C2_D2}), it can be derived that
\begin{align}
\left[ \begin{matrix}
\Phi_1 - \lambda I & \tilde E \\
C_2 & {H}_{\tau}^f
\end{matrix}
\right]
= \left[ \begin{matrix}
\Phi - \lambda I & \tilde E \\
C \Phi^{\tau} & {H}_{\tau}^f
\end{matrix}
\right] 
\left[ \begin{matrix}
I & 0 \\
-\Pi C \Phi^{\tau} & I
\end{matrix}
\right].
\end{align}
With Assumption \ref{ass:fault_rank}, if $( \Phi, \tilde E, C \Phi^{\tau}, {H}_{\tau}^f )$ has no unstable invariant zeros, it follows that
\begin{equation*}
\begin{aligned}
&\text{rank} \left[ \begin{matrix}
\Phi_1 - \lambda I & \tilde E \\
C_2 & {H}_{\tau}^f
\end{matrix}
\right]  
= \text{rank}  \left[ \begin{matrix}
\Phi - \lambda I & \tilde E \\
C\Phi^{\tau} & {H}_{\tau}^f
\end{matrix}
\right] 
= n + n_f
\end{aligned}
\end{equation*}
for $\left| \lambda \right| \ge 1$, which implies (\ref{eq:unobsv_mode}).
The proof for observability of $( \Phi_1, C_2 )$ is similar, thus is omitted. 

\section*{Proof of $\mathcal{K}_L = \mathcal{G}_L + \mathcal{M}_L \mathcal{J}_L$}\label{app:equal_KL}
Since $\mathcal{K}_L$, $\mathcal{G}_L$, $\mathcal{M}_L$, and $\mathcal{J}_L$ are block-Toeplitz matrices defined in (\ref{eq:GL_JL}), we can prove 
$\mathcal{K}_L = \mathcal{G}_L + \mathcal{M}_L \mathcal{J}_L$ by proving 
\begin{equation}\label{eq:aimproof}
K_i = G_i + \sum_{j=0}^{i} M_{i-j} J_j 
\end{equation}
where $K_i$, $G_i$, $M_i$ and $J_i$ are the MPs that construct $\mathcal{K}_L$, $\mathcal{G}_L$, $\mathcal{M}_L$, and $\mathcal{J}_L$, respectively, as described in (\ref{eq:RL_QL_Markov}).
According to (\ref{eq:GL_JL}), these MPs are
\begin{equation}\label{eq:MP_app}
\begin{aligned}
& K_i = \left\{ \begin{array}{ll}
D_1 & i=0 \\
C_1 (\Phi_1 - K_r C_2)^{i-1} (B_1 + K_r D_2)  & i>0
\end{array} \right. , \\
& G_i = \left\{ \begin{array}{ll}
D_1 & i=0 \\
C_1 \Phi_1^{i-1} B_1 & i>0
\end{array} \right. ,\;
J_i = \left\{ \begin{array}{ll}
D_2 & i=0 \\
-C_2 \Phi_1^{i-1} B_1 & i>0
\end{array} \right. ,\\
& M_i = \left\{ \begin{array}{ll}
0 & i=0 \\
C_1 (\Phi_1 - K_r C_2)^{i-1} K_r & i>0
\end{array} \right. .
\end{aligned}
\end{equation}

By using (\ref{eq:MP_app}), we can prove (\ref{eq:aimproof}) as follows:
\begin{equation*}
\begin{aligned}
K_i &= C_1 (\Phi_1 - K_r C_2)^{i-1} B_1 + M_i J_0 \\
&= C_1 (\Phi_1 - K_r C_2)^{i-2} (\Phi_1 B_1 - K_r C_2 B_1) + M_i J_0 \\
&= C_1 (\Phi_1 - K_r C_2)^{i-2} \Phi_1 B_1 + M_{i-1} J_1 + M_i J_0 \\
&= C_1 (\Phi_1 - K_r C_2)^{i-3} \Phi_1^2 B_1 + \sum_{j=0}^{2} M_{i-j} J_j \\
& \cdots \\
&= C_1 (\Phi_1 - K_r C_2) \Phi_1^{i-2} B_1 + \sum_{j=0}^{i-2} M_{i-j} J_j \\
&= G_i + \sum_{j=0}^{i-1} M_{i-j} J_j + M_0 J_i.
\end{aligned}
\end{equation*}

\section*{Proof of Lemma \ref{lem:alpha}}\label{app:alpha}
Proof of (\romannumeral1): Suppose that $\left[ \begin{smallmatrix}
\alpha \\ \Pi
\end{smallmatrix} \right]$ is singular, then there exist nonzero vectors $\beta_1 \in \mathbb{R}^{n_y - n_f}$ and $\beta_2 \in \mathbb{R}^{n_f}$ such that $\beta_1^\text{T} \alpha + \beta_2^\text{T} \Pi = 0$. Under this condition, we have 
$ (\beta_1^\text{T} \alpha + \beta_2^\text{T} \Pi) H_\tau^f = 0
$.
The above two equations imply $\beta_1 = 0$ and $\beta_2 = 0$ because 
$\alpha H_\tau^f = 0$ (see Step (\romannumeral2) of Algorithm B) and $\Pi H_\tau^f = I$. The contradiction with the nonzero assumption about $\beta_1$ and $\beta_2$ proves (\romannumeral1).

Proof of (\romannumeral2): Using $\Pi H_\tau^f = I$, (\ref{eq:Ji}) and (\ref{eq:Qi}), it is straightforward but tedious to verify that $\Pi J_i = 0$ and $\Pi Q_i = 0$ for $i \geq 0$. Then we have
$\bar \Pi \mathcal{H}_Q = 0$, where $\mathcal{H}_Q$ is defined in (\ref{eq:HRL}) and $\bar \Pi = \text{diag}(\Pi, \Pi, \cdots, \Pi)$ with $l$ diagonal blocks. Note that we have 
$\text{col}(\mathcal{\hat O}_Q) = \text{col}(\mathcal{\hat H}_Q) \subseteq \text{col}(\mathcal{H}_Q) $ 
for the rank-reduced block-Hankel matrix $\mathcal{\hat H}_Q$ and the corresponding observability matrix $\mathcal{\hat O}_Q$, where $\text{col}(X)$ represents the column space of a matrix $X$. Then $\mathcal{\hat O}_Q$ satisfies $\bar \Pi \mathcal{\hat O}_Q = 0$. Consequently, $\hat C_2$ obtained in (\ref{eq:C2_hat}) ensures the condition (\romannumeral2).

\section*{Proof of Theorem \ref{thm:ARE}}\label{app:ARE}
It is well known that the ARE (\ref{eq:ARE}) has a unique stabilizing solution if and only if $(\hat{\Phi}_1, \bar C_2)$ is detectable and $(F_s, Q_s^{\frac{1}{2}})$ is controllable on the unit circle, see Appendix E of \cite{Kailath2000}, where 
\begin{subequations}\label{eq:FsQs}
	\begin{align}
		F_s &= \hat \Phi_1 - \hat B_1 \Sigma_e \bar D_2^\text{T} \left( \bar D_2 \Sigma_e \bar D_2^\text{T} \right)^{-1} \bar C_2, \\
		Q_s^{\frac{1}{2}} &= \hat B_1 \Sigma_e^{\frac{1}{2}} - \hat B_1 \Sigma_e \bar D_2^\text{T} \left( \bar D_2 \Sigma_e \bar D_2^\text{T} \right)^{-1} \bar D_2 \Sigma_e^{\frac{1}{2}}.
	\end{align}
\end{subequations}
Using Lemma \ref{lem:alpha}, the detectability of $(\hat{\Phi}_1, \bar C_2)$ is given by 
\begin{equation}
\begin{aligned}
\text{rank} \left[ \begin{matrix}
\hat \Phi_1 - \lambda I \\
\bar C_2
\end{matrix} \right] &
= \text{rank} \left[ \begin{matrix}
I & 0 \\
0 & \alpha \\
0 & \Pi \\
\end{matrix} \right]
\left[ \begin{matrix}
\hat \Phi_1 - \lambda I \\
\hat C_2
\end{matrix} \right] \\
&= \text{rank} 
\left[ \begin{matrix}
\hat \Phi_1 - \lambda I \\
\hat C_2
\end{matrix} \right]
= n, \text{ for } |\lambda| \geq 1,
\end{aligned}
\end{equation}
which proves (\ref{eq:obsvb}).

Let $\lambda_1$ be an uncontrollable mode of $(F_s, Q_s^{\frac{1}{2}})$. This is equivalent to the existence of a nonzero row vector $\nu$ such that 
\begin{equation*}
\nu \left[\begin{matrix}
F_s - \lambda_1 I & Q_s^{\frac{1}{2}}
\end{matrix}\right] = 0.
\end{equation*}
From (\ref{eq:FsQs}), the above equation can be rewritten as 
\begin{equation*}
\nu_1 \left[ \begin{matrix}
\hat \Phi_1 - \lambda_1 I & \hat B_1 \\
\hat C_2 & \hat D_2
\end{matrix} \right]
\left[ \begin{matrix}
I & 0 \\
0 & \Sigma_e^{\frac{1}{2}}
\end{matrix} \right] = 0
\end{equation*}
with $\nu_1 = \nu \left[ \begin{matrix}
I & -\hat B_1 \Sigma_e \bar D_2^\text{T} \left( \bar D_2 \Sigma_e \bar D_2^\text{T} \right)^{-1}
\alpha
\end{matrix} \right]$.
Therefore, the controllability of $(F_s, Q_s^{\frac{1}{2}})$ on the unite circle is equivalent to (\ref{eq:contrl}).



%



\ifCLASSOPTIONcaptionsoff
  \newpage
\fi



\bibliographystyle{IEEEtran}
\bibliography{ID_fest_filter}

%
%
%

%


\vfill






\end{document}